\documentclass[a4paper,12pt]{article}
\usepackage[utf8]{inputenc}
\usepackage[english]{babel}
\usepackage[margin=1.2in]{geometry}
\usepackage{amsthm,amsmath,amsfonts,amssymb}
\usepackage[authoryear]{natbib}
\newcommand{\new}[1]{\textcolor{black}{#1}}
\usepackage{graphicx}%% uncomment this for including figures

\RequirePackage{enumitem}

\usepackage{xcolor}
\usepackage[ruled, vlined]{algorithm2e}
\usepackage{multicol}
\usepackage{multirow}
\usepackage[normalem]{ulem}
\usepackage[colorlinks,citecolor=blue,urlcolor=blue]{hyperref}%% uncomment this for coloring bibliography citations and linked URLs
\newcommand{\eqname}[1]{\tag*{(#1)}}
\newtheorem{properties}{Properties}
\newtheorem{remark}{Remark}
\newcommand{\pr}{\mathbb{P}}
\newcommand{\E}{\mathbb{E}}
\newcommand{\btheta}{\boldsymbol{\theta}}
\newcommand{\btau}{\boldsymbol{\tau}}
\newcommand{\balpha}{\boldsymbol{\alpha}}
\newcommand{\bpi}{\boldsymbol{\pi}}
\newcommand{\bZ}{\boldsymbol{Z}}
\newcommand{\bdelta}{\boldsymbol{\delta}}
\newcommand{\pen}{\mbox{pen}}
\newcommand{\argmax}{\mbox{argmax}}
\newcommand{\obs}{\mathbf{X}}
\newcommand{\lat}{\mathbf{Z}}
\newcommand{\Xm}{X^{m}}
\newcommand{\Zm}{Z^{m}}
\newcommand{\pim}{\pi^{m}}
\newcommand{\alpham}{\alpha^{m}}
\newcommand{\deltam}{\delta_{m}}
\newcommand{\taum}{\tau^{m}}

\newcommand{\balpham}{\balpha^{m}}
\newcommand{\bpim}{\bpi^{m}}
\newcommand{\Sc}{\text{Sc}}
\newcommand{\con}{\alpha}
\newcommand{\Nb}{\mbox{NP}}

\newcommand{\dens}{\delta}
\newcommand{\nm}{n_m}
\newcommand{\colSBM}{colSBM}
\newcommand{\ICL}{\text{ICL}}
\newcommand{\BICL}{\text{BIC-L}}
\newcommand{\eBICL}{\text{BIC-L}}
\newcommand{\iidcolSBM}{iid\text{-}\colSBM}
\newcommand{\denscolSBM}{\dens\text{-}\colSBM}
\newcommand{\picolSBM}{\pi\text{-}\colSBM}
\newcommand{\denspicolSBM}{\dens\pi\text{-}\colSBM}
\newcommand{\sepSBM}{sep\text{-}SBM}
\newcommand{\ARI}{\text{ARI}}
\newcommand{\Qcal}{\mathcal{Q}}
\newcommand{\Qbf}{\mathbf{Q}}
\newcommand{\Mcal}{\mathcal{M}}
\newcommand{\Acal}{\mathcal{A}}
\newcommand{\FSBM}{\Fcal\mbox{-SBM}}
\newcommand{\SBM}{\mbox{-SBM}}
\newcommand{\Scal}{\mathcal{S}}
\newcommand{\Rcal}{\mathcal{R}}
\newcommand{\Gcal}{\mathcal{G}}
\newcommand{\Fcal}{\mathcal{F}}

\usepackage{authblk}
\title{Learning common structures in a collection of networks. An application to food webs}

\author[1]{Saint-Clair Chabert-Liddell}
\author[1]{Pierre Barbillon}
\author[1]{Sophie Donnet}

\affil[1]{Université Paris-Saclay, AgroParisTech, INRAE, UMR MIA Paris-Saclay, 75005, Paris, France}
\date{}

\begin{document}
\maketitle
 
\begin{abstract}
Let   a collection of networks represent  interactions within several (social or ecological) systems. We pursue two objectives:  \new{identifying similarities in the topological structures that are held in common between the networks and  clustering the collection into sub-collections of structurally homogeneous networks. }%\sout{identifying similarities between the topological structures of the networks or  clustering the networks according to the similarities in their structures.}
We tackle  these two questions with a  probabilistic model based approach. We propose an extension of the Stochastic Block Model (SBM) adapted to the joint  modeling of a collection of networks. The networks in the collection are assumed to be independent realizations of SBMs. The common connectivity structure  is imposed through the equality of some parameters. %, possibly up to the block proportions and/or a density factor.

The model parameters are estimated with a variational Expectation-Maximization (EM) algorithm. We derive an ad-hoc penalized likelihood  criterion to select the number of blocks and to assess the adequacy of the consensus found between the structures of the different networks. This same criterion can also be used to cluster networks on the basis of their connectivity structure. It thus provides a partition of the collection into subsets of structurally homogeneous networks.

The relevance of our proposition is assessed on two collections of ecological networks. First, an application to three stream food webs reveals the homogeneity of their structures and the correspondence between groups of species in different ecosystems playing equivalent ecological roles.
Moreover, the joint analysis   allows a finer analysis of the structure of smaller networks. Second, we cluster $67$ food webs according to their connectivity structures and demonstrate that five mesoscale structures are sufficient to describe this collection.
\end{abstract}

% \begin{keyword}
% \kwd{Stochastic Block Model}
% \kwd{Clustering}
% \kwd{Networks}
% \kwd{Latent variable models}
% \kwd{Ecology}
% \end{keyword}

%========================================================
%====================================== MAIN TEXT 

\section{Introduction}

\paragraph*{Context}
The last few years have seen an increase in the number of interaction networks collected, as networks are popular tools for representing the functioning of a social or ecological system.
For a long time the statistical analysis of network data has focused on analyzing a single network at a time. This can be performed either by looking at local or global topological features, or by setting a probabilistic model inferred from the network \citep{kolaczykstatistical}.
When several networks describing the same kind of interactions are available, a natural question is to assess to what extent they are similar or different.
As network data are complex by nature, this comparison of different networks is not an easy task and has mainly focused on comparing statistical topological features on the local, global or mesoscale levels. \new{These comparison metrics depend on whether the networks are defined on the same set of nodes or on different sets of nodes . A survey for the former case is dealt with in \citet{donnat2018} and another survey for both cases is done in \citet{wills2020metrics}.} %\sout{reviews a review of} \sout{graph} distances \new{between graphs involving the same nodes}].%{donnat2018, wils2020metrics}.

% dire ce qu'on entend par collection de réseaux noeuds different mais meme type d'interaction

% interet de l'inference jointe a presenter

In this paper, we consider networks with no node correspondence and no link between networks
%\old{We assume that the considered networks have no nodes in common  and that the nodes of different networks are not linked}
as it may be the case in multilayer networks \citep{kivela2014multilayer}. Furthermore, the networks are assumed to represent interactions of the same type (directed or not) and with the same valuation (binary, discrete or continuous).
A set of such networks constitutes what we call in this paper a collection of networks, although some authors may use this terminology in a different meaning.

When observing such a collection, we aim to determine if the respective structures of the networks are similar. This paper focuses on the mesoscale structure of the networks by assuming that the nodes can be grouped into blocks on the basis of  their connectivity pattern \citep{white1976social}.
A classical tool to infer such a  mesoscale structure of a single network is the Stochastic Block Model \citep[SBM]{holland1983stochastic, snijders1997estimation}. In the SBM, a latent variable is associated with each node giving its group/block membership. Nodes belonging to the same block share the same connectivity pattern. The SBM has easily interpretable parameters and
its framework allows multiple extensions such as modeling the interactions with various distributions
%such as allowing interaction to take discrete or continuous value
\citep{mariadassou2010uncovering}.  \new{The block memberships are not known a priori, they are recovered a posteriori by the inference algorithm.}

\new{ In social (resp. ecological) networks, individuals (resp. species) with the same block membership play the same social/ecological role in its system \citep{boorman1976, luczkovich2003}. In food webs, species playing the same ecological role are said to be ecologically equivalent \citep[see][for a review of species role concepts in food webs]{cirtwill2018review}.  When analysing the  roles in food webs, \citet{luczkovich2003}  use the notion of regular equivalence to define trophic role. Two species are said to be regularly equivalent if they feed on equivalent species and are preyed on by equivalent species. This notion of regular equivalence is a relaxation of structural equivalence which imposes that structurally equivalent species have exactly the same trophic relations in the food web.
  In practice, \citet{luczkovich2003} find that species are grouped into blocks by trophic level and some separation might occur based on trophic chains.
  Other papers lead to a similar interpretation of the blocks for stochastic equivalence when fitting SBMs on food webs. It is also noticed that communities (blocks of species preying on each other) are  unusual \citep{allesina2009food,sander2015can}.
Stochastic equivalence has the advantage of taking into account the noisy aspects of the observed networks.
In addition, SBM, as a probabilistic generative model, provides the modeler with a unified framework for model selection, link prediction, simulation, and modeling extension, for example, for a collection of networks.
%considering model selection criterion, the SBM might lack the statistical power to find detailed block on each level and as such might provide less refined blocks than deterministic tools using regular equivalence.
}

Inferring independently an SBM for each network and  comparing them may be misleading. Indeed, a given network may have several possible grouping of the nodes into blocks  \citep{peel2017ground} that are equally likely \citep{peixoto2014hierarchical}.
Furthermore, the observation of a network may be noisy \citep{guimera2009missing}, especially for ecological networks, the sampling of which is known to be incomplete \citep{rivera2012effects}.

%\textcolor{red}{Sophie : je comprends pas trop ce paragraphe?}

\paragraph*{Our contribution}
Thus, we propose to  jointly model a collection of networks by extending the SBM. We assume that the networks are independent realizations of SBMs sharing  common parameters. The natural and interesting consequence is the correspondence between the blocks of the different networks.
The proposed model called $\colSBM$ comes with a few variants. The simplest model  assumes that the parameters of the SBMs are identical
leading to a collection of i.i.d. networks.
As this assumption might be too restrictive for real networks, we introduce two relaxations on this assumption.
The first one is to allow the distribution of the block memberships to vary between networks and even to allow some networks to not populate certain blocks. This enables to model a collection of networks where the structure of certain networks is encompassed in the structure of other networks.
The second relaxation allows networks to have the same structure up to a density parameter. This is particularly useful to model networks with different sampling efforts, since it has a direct impact on the density of ecological networks \citep{bluthgen2006measuring}.

The inference of the block memberships, the model parameters and the model selection are done through an ad-hoc version of classic tools when inferring SBM, namely a  Variational EM algorithm for the inference and an adaptation of the integrated classification likelihood (ICL) criterion for the model selection \citep{daudin2008mixture}.

The interest of our  $\colSBM$ model is two-folds. The first one is to find a common connectivity pattern which explains the structure of the different networks in the collection and to assess via model selection whether these structures are a reasonable fit for the collection. As a  by-product, it allows a fine analysis of the role structure of the different nodes in the networks.
%\old{In social/ecological networks, individuals/species with the same block membership play the same social/ecological role in its system \citep{boorman1976, allesina2009food}.}
By sharing the blocks between the networks, $\colSBM$ allows to recover sets of nodes which play the same sociological/ecological roles in different networks.
The second one is to provide %subsets of networks that share common connectivity patterns.%,giving us such
a partition of the collection of networks into sub-collections of structurally homogeneous networks\footnote{\new{For the sake of clarity, we chose to use the terminology sub-collection, cluster and clustering for partitioning a set of networks while we use blocks or groups (and grouping) when referring to grouping nodes of a network into blocks.}}.
Both aspects have some practical implications in ecology \citep{Ohlsson2020, Michalska-Smith2019}.

As a side effect, by modeling these networks together, provided that the networks have common connectivity patterns, we can use the information of certain networks to recover noisy information from other networks by improving the prediction of missing links \citep{clauset2008hierarchical}. Hence $\colSBM$ has a stabilizing effects on the grouping of the nodes into blocks and might give a block membership that is closer to the one of the full real network than just a single SBM as this will be shown in the numerical studies and application.

%\STC{}{Il faut revoir la partie Related Works et Retravailler l'outline suivant le plan final.}
\paragraph*{Related work}
Since the SBM is a very flexible model, it has already  been adapted to multilayer networks. To name a few, \citet{matias2017statistical} model a collection of networks along a time gradient, the connectivity structure varies from time to time but they integrate a sparsity parameter, which is similar to our density parameter in the binary case.%, for identifiability issue they make block memberships vary with between time stamps but sone of the connectivity parameter has to stay constant across time.%, similar to the work of \citet{yang2011detecting}.
When dealing with networks with no common nodes, \citet{chabert2021multilevel} deal with multilevel networks where the networks are linked by a hierarchical relation between the nodes of the different levels. Within the SBM framework, the closest work  to ours is the strata multilayer SBM  \citep{stanley2016clustering}, in that it looks for both common connectivity patterns and network clustering. However, it does not consider a collection of networks but a multiplex network where all the networks share the same nodes.
% It considers a multiplex network with common nodes on each network, but the networks are classified into groups that share the same connection patterns.

Most contributions about collections of networks rely on some node correspondence between the networks. Recently, motivated by the analysis of fMRI data a few works extend the SBM to model population of networks \citep{paul2018, pavlovic2020}. \citet{le2018} make the assumption that the networks of the collection are noisy realizations of the true network, while \citet{reyes2016stochastic} use in a Bayesian framework a hierarchical SBM to model the collection. \new{\citet{durante2017nonparametric} propose a mixture of latent space models and }\citet{signorelli2020model}  a mixture of network models which is not restricted to the SBM.

\new{Dealing with networks with no node correspondence, \citet{faust2002} compare networks involving different species and interaction types using the parameters of exponential random graph models (ERGMs or $p^{*}$ models). More recently, \citet{yin2022} propose a mixture of ERGMs to model the generative process of a collection of networks. ERGMs allow testing the significance of selected local interaction patterns that convey ecological or sociological meaning. Compared to $\colSBM$, they do not group nodes into blocks and as such do not provide role equivalence between nodes of different networks. }
The contributions dealing with networks with no node correspondence also  include  a hierarchical mixed membership SBM, using a common Bayesian prior on the connectivity parameter of the different networks \citep{sweet2014hierarchical}.
Finally on partitioning a collection of networks, \citet{mukherjee2017clustering} use graph moments (they also propose to fit a mixture of graphon when having access to node correspondence  between the networks in the collection and then to make a spectral clustering on the distance matrix between networks)\new{, while \citet{sweet2019} use graph kernel methods on networks with nodes label to estimate independently a feature vector for each network. Both of these contributions rely on clustering those feature vectors, and as such do not provide any estimate on the joint structure of the collection.}

%\citep{, durante2017nonparametric, }
%\new{Comparaison p*(ERGM) \citep{faust2002}, ergm mixture : \citep{yin2022}, classif kernel \citep{sweet2019}, mixture : \citep{durante2017nonparametric})}

\paragraph*{Outline}
Section  \ref{coll:sec:sbm} recalls the definition of the Stochastic Block Model on a single network. We  motivate our new approach by inferring it independently on a collection of food webs. Then in Section \ref{coll:sec:model}, we present the various variants of the $\colSBM$. The likelihood expression is provided in Section \ref{coll:sec:properties}, together with some identifiability  conditions. %in Section \ref{coll:sec:properties}.
We develop the methodology for the parameter estimation and model selection in Section \ref{coll:sec:inference}, while Section \ref{coll:sec:partition} deals with network clustering. \new{The details on the clustering procedure are postponed to Appendix \ref{coll:ap:sec:partition}.}. % where we explain how the model selection procedure developed previously could be used for this purpose.
We finally propose with two applications on food webs in Section \ref{coll:sec:foodwebs}. First, we compare the structures of $3$ networks and show the information transfer between these networks. Second, we seek a partition of a  collection of $67$ networks. The technical details and  numerical studies which demonstrates the efficiency of our inference procedure and the pertinence of our model selection criterion are left in Supplementary Material \citep{supplementary}. %.   \ref{coll:sec:identifiability_proof} and \ref{coll:sec:detail_icl}.

\section{Data motivation and the stochastic block model}\label{coll:sec:sbm}

Consider a collection of $M$ independent networks where  each network indexed by $m$   involves its own $\nm$ nodes.
The networks are encoded into their adjacency matrices $(\Xm)_{m \in \{1, \dots, M\}}$ such that: $\forall  m \in \{1, \dots, M\}$, $\forall (i \neq j) \in \{ 1, \dots, \nm\}^2$,
\begin{equation*}
   \left\{
\begin{array}{ccl}
 X^{m}_{ij} &=  0 & \mbox{ if no interaction is observed between species $i$ and $j$ of network $m$}\\
 X^{m}_{ij} &\neq  0  & \mbox{ otherwise}.
\end{array}
\right.
\end{equation*}
If the networks represent binary interactions then $X^{m}_{ij} \in \mathcal{K} =  \{0,1\}$, $\forall (m,i,j)$; if the interactions are weighted such as counts, then $X^{m}_{ij} \in \mathcal{K} = \mathbb{N}$. Moreover, all the networks  encompass the same type of  interactions (binary, count\dots) and no self-interaction is considered. Besides,  for the sake of simplicity,  we assume that all the networks are directed.  The extension  to undirected networks i.e.  such that $X^{m}_{ij} = X^{m}_{ji}$ for any $i \neq j$  is straightforward.
$\obs = (X^1, \dots, X^{M})$ denotes the collection of adjacency matrices.

\paragraph*{A first ecological example: three stream food webs}\label{coll:ex stream}
As a first example, we  consider
 the collection of three  stream food webs   from  \cite{thompson2003impacts}.  The three networks collected respectively in Martins (Maine USA), Cooper and Herlzier (North-Carolina, USA) involve respectively $105$, $58$ and  $71$ species resulting in  $343$, $126$ and $148$ binary edges  respectively.
 Classically, the food web  edges represent directed trophic links showing the energy flow ie. $X^{m}_{ij} = 1$ if species $j$ preys on species $i$,  with no reciprocal interactions. When aiming at unraveling the structure of these networks % in order to  have a general view of the trophic relations,
 the Stochastic Block Model (SBM)   is an interesting tool which  has proven its high flexibility by encompassing a large variety of structures \citep[see][for the particular case of food webs]{allesina2009food}. %The SBM consists in clustering together nodes of a network which have the same patterns of connection.
 When dealing with three networks, the standard strategy that we describe below, is to fit separately one SBM per network.
 \\

\paragraph*{Separate SBM (sepSBM)}

%Standard strategies would imply studying the structure of each network separately. To do so, the Stochastic Block Model (SBM) \cite{} has proven its ability to deciphere  the structure that underlies a network, may it represent binary or weighted interactions.
The SBM introduces blocks of nodes and assumes that the interaction between two nodes is driven by the blocks the nodes belong to.
More precisely, for network $m$, let the  $\nm$ nodes be divided into $Q_m$ blocks.  Let $\Zm = (\Zm_{1}, \dots ,\Zm_{\nm})$ be independent latent random variables  such that  $\Zm_{i} = q$  if node $i$ of network $m$ belongs to  block $q$ with $q \in \{1, \dots, Q_m\}$ and
\begin{equation}\label{coll:eq:SBM1}
   P(\Zm_{i}=q) =\pim_{q}
\end{equation}
where   $\pim_q > 0$ and $ \sum_{q = 1}^{Q_m} \pim_{q} = 1$.
Given the latent variables $\Zm$, the $\Xm_{ij}$'s are assumed to be independent and distributed as
%The edges of the network are then distributed depending on the clusters of its nodes. For nodes $i < j$ in network $m$:
\begin{equation}\label{coll:eq:SBM2}
  \Xm_{ij} | \Zm_{i}=q, \Zm_{j} = r  \sim \Fcal(\cdot; \alpha^{m}_{qr}),
\end{equation}
where  %We will assume in the following that , and takes discrete one dimensional values. Mainly, the interactions could be:
$\Fcal$ is  referred to as the emission distribution. $\Fcal$ is  chosen to be the Bernoulli distribution for binary interactions, and the Poisson distribution for weighted interactions such as counts.   Let $f$ be the density of the emission distribution, then:
\begin{equation}\label{coll:eq:emission_dens}
 \log f(\Xm_{ij};\con^m_{qr}) = \left\{
 \begin{array}{ll}
\Xm_{ij}  \log\left(\con^m_{qr}\right)  + (1-\Xm_{ij}) \log\left(1- \con^m_{qr}\right) & \mbox{for Bernoulli emission}\\
- \con^m_{qr}+ \Xm_{ij}  \log\left(\con^m_{qr}\right)  - \log( X_{ij}!) & \mbox{for Poisson emission}\\
\end{array}
\right.
.
\end{equation}
%
%
% Then
%   \begin{equation*}
%     f(\Xm_{ij} |  \Zm_{i}=q, \Zm_{j} = r; \alpha^{m}_{qr}) = {\alpha^m_{qr}}^{\Xm_{ij}}(1-\alpha^m_{qr})^{1-\Xm_{ij}}
%   \end{equation*}
% for binary interactions ($\Xm_{ij} \in \{0,1\}$);
% or the  Poisson distribution with density
%   \begin{equation*}
%     f(\Xm_{ij} | \Zm_{i} = q, \Zm_{j} = r; \alpha^{m}_{qr}) = e^{-\alpha^m_{qr}}\frac{{(\alpha^m_{qr})}^{\Xm_{ij}}}{\Xm_{ij}!}.
%   \end{equation*}
% for weighted interactions.
%\PB{}{est ce que ça ne serait pas plus simple de dire, on suppose un modèle sbm pour chaque m et rappeler le sbm à m fixé, je pense que c'est aussi important de préciser que le sbm a des paramètres propres à chaque réseau dans cette premiere présentation  que que tu vas ensuite proposer des hypothèses sur ces paramètres}
%  \PB{}{pourquoi ne pas utiliser la notation $Z_i=q$ ?}
%\end{itemize}
 %Extension to directed networks or count data will be left in appendix when not straightforward.
Equations \eqref{coll:eq:SBM1}, \eqref{coll:eq:SBM2} and   \eqref{coll:eq:emission_dens} define the SBM model and we will now use   the following short notation:
\begin{equation}\label{coll:SBMmodel}
 \Xm \sim \Fcal\mbox{-SBM}_{\nm}(Q_m, \bpi^m,\balpha^m).
 \eqname{$sep\text{-}SBM$}
\end{equation}
where $\Fcal$ encodes the emission distribution, $\nm$ is the number of nodes, $Q_m$ is the number of blocks  in network $m$, and $\bpim = (\pim_q)_{q =  1, \dots, Q_m}$ is the vector of  their proportions. The $Q_m \times Q_m$ matrix $\balpham = \left(\alpham_{qr}\right)_{q, r = 1, \dots, Q_m}$ denotes  the connection parameters i.e. the parameters of the emission distribution. %(in the case of directed interactions).
Moreover, $\alpha^m_{qr}\in \Acal_\Fcal$ where $\Acal_\Fcal = (0,1)$ (resp. $\Acal_\Fcal=\mathbb{R}^{*+}$)  for the Bernoulli (resp. Poisson) emission distribution.
%$\theta^m = (\pim, \alpham)$.
In the $sep\text{-}SBM$ model, each network $m$ is assumed to follow a SBM with its own parameters $(\bpim, \balpham)$.

\paragraph*{\new{First ecological example}: three stream food webs}\label{coll:sepSBM:stream}
We fit the \textit{sep-SBM} on the $3$ stream food webs, respectively referred to as  Martins, Cooper and Herlzier.  To do so, we use the \texttt{sbm} R-package \citep{sbmpackage,blockmodelspackage} on each network,  which implements a variational version of the EM algorithm  to estimate the parameters   and selects the number  of  blocks $Q_m$ using a penalized likelihood criterion ICL. These inference tools (variational EM and ICL) will be recalled hereafter.

\begin{figure}[!t]
{\centering{
    \includegraphics[width = .8\hsize]{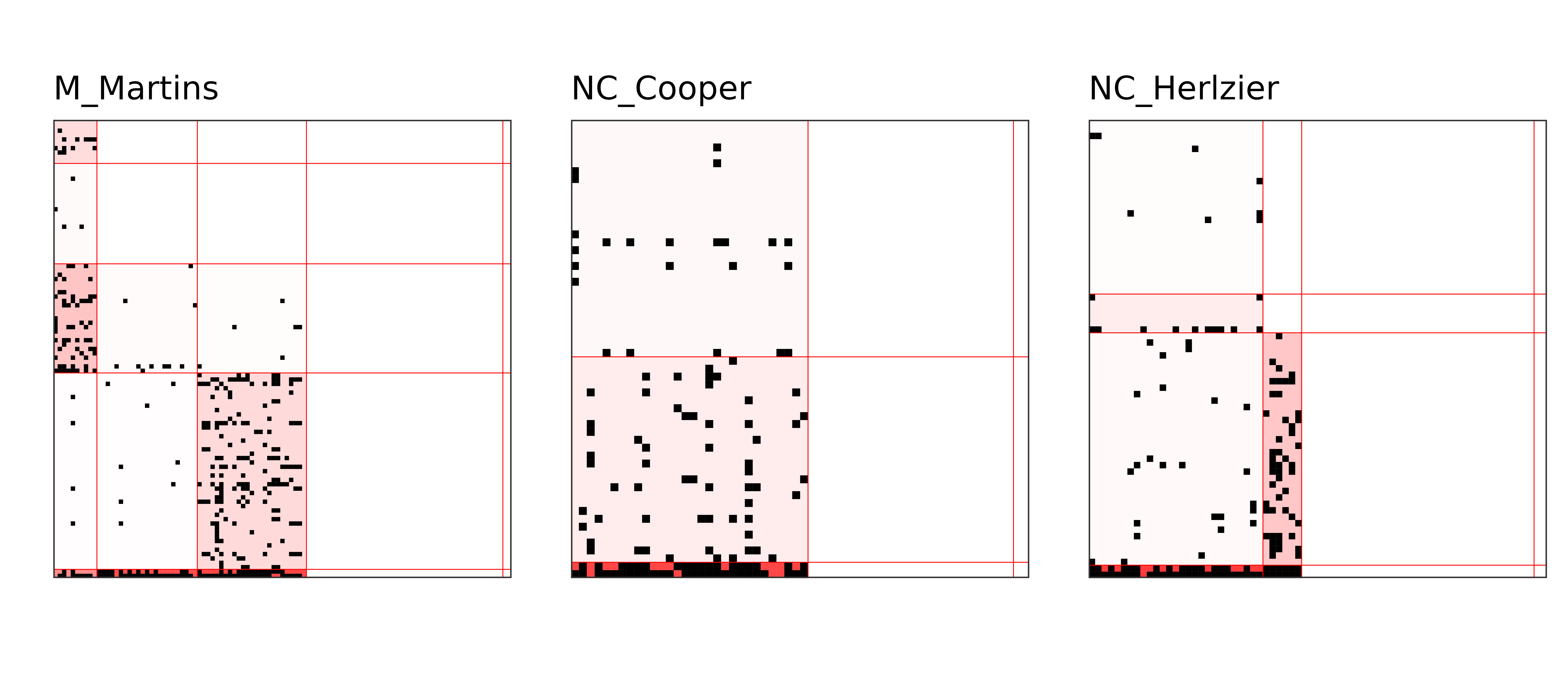}}
  \caption{Matricial view of $3$ stream food webs. The species are reordered by blocks and blocks are ordered by expected out-degrees to emulate the trophic levels (bottom to top and right to left).  The blocks have been obtained by fitting a SBM on each network separately. \new{The shades of red depict the connectivity parameters $\widehat{\balpha}^{m}$. }\label{coll:fig:3net_sbm}}}
\end{figure}

We  obtain respectively $\widehat{Q}_1 = 5$ blocks for Martins, $\widehat{Q}_2= 3$ blocks for Cooper and $\widehat{Q}_3 = 4$ blocks for Herlzier. The adjacency matrices of the food webs reordered by block membership  are plotted in Figure \ref{coll:fig:3net_sbm}. \new{The two bottom blocks of each food web is composed   of basal species (species not feeding on other species). For Cooper, the higher trophic levels are grouped together in the same block: the lack of statistical power does not allow to divide the nodes into more blocks.}  For Herlzier the higher trophic level is separated into $2$ blocks mainly determined on how much they prey on the less preyed basal block. Martins has a separation into $3$ blocks, the third one is a medium trophic level, which preys on basal species and is highly preyed on by species of the first block.
The first two blocks are made up of higher trophic level species, with the last two blocks being much less connected than the first.

 As can be seen in Figure \ref{coll:fig:3net_sbm}, the connectivity structures of these three networks seem to have a lot of similarities. To explore further this aspect,  Section \ref{coll:sec:model} is dedicated to the presentation of several  $\colSBM$  models  assuming common structures among the networks of a given collection.

% \new{Introduire l'appli rmangal en disant que si on considère plus de réseaux que 3, on peut supposer que tous n'ont pas la même structure, et que l'on souhaite trouver un ensemble de structures qui décrit bien la collection ce qui nous permet de regrouper des réseaux semblables.}

%  \PBs{pas fan de dire a second ecological example c'est le meme non ? Il faudrait Study of the whole collection of foodwebs ou of a large collection of ...}
\new{\paragraph*{A second ecological example: $67$ predation networks}
\label{coll:sepSBM:predation} The three previously presented food webs were extracted  from a larger collection which involves more  networks collected in similar ecosystems. When willing to analyze a larger collection, with more heterogeneous conditions, one cannot expect to find a structure that will fit all networks well, but it could fit a sub-collection of networks.
 To illustrate this, we consider a collection of $67$ predation networks which are all directed networks with more than $30$ species, from the Mangal database \citep{rmangal}. They are issued from $33$ datasets each containing from  $1$ to  $10$ networks. The number of species ranges from $31$ to $106$ ($3395$ in total) by networks; the networks   have density ranging from $.01$ to $.32$ ($14934$ total predation links). Fitting $\sepSBM$ on this collection leads to networks having  between $2$ to $8$ blocks. The number of blocks containing only basal species varies from $0$ to $3$ betwwen the 67 networks which leads to quite contrasted structures. A method to cluster the networks of this collection will be developed in Section \ref{coll:sec:partition}.}
% introducing different types of  consensus between the structures of a collection of  networks.
% . % To do so, we assume that each network $m$ has a set  of latent variables $\lat = (Z^1, \dots, \Zm)$ which take values in a space $\mathcal{Q} = \{1, \dots, Q\}$ that is common for all the networks.
% Given this framework we exhibit several probabilistic models introducing different levels of consensus between the networks of the collection.

% \PB{}{Tu ne considère pas comme une alternative possible chaque réseau avec ses propres paramètres du moins pour la comparaison ??? ou en fait c'est plus vicieux et c'est un sous-cas de pi col avec des blocs disjoints}
\section{Joint modeling of  a collection of networks}\label{coll:sec:model}

We now present a set of probabilistic models designed to introduce structure consensus into a collection of networks of interest. For ease of notation, we develop the models for directed networks; extensions to the undirected cases are straightforward. Note that the networks $(\Xm)_{m=1, \dots, M}$  are always assumed to be independent random objects.
%For all $m \in \Mcal$, the  are independent
A summary of the various models is provided in Table \ref{coll:tab:summarymodels}, from the most to the less constrained model

\subsection{A collection of i.i.d. SBM}
% ($iid{\text -}\colSBM$)
The first model we propose is the most constrained one and assumes that the networks are independent realizations of the same $Q$-block  SBM model with identical parameters. The so-called $ \iidcolSBM$
states that:

\begin{equation}\label{coll:mod:iid}
 \Xm \sim \FSBM_{n_m}(Q, \bpi,\balpha), \quad \quad  \quad \forall m=1, \dots M,
 \eqname{$\iidcolSBM$}
\end{equation}
where   $\forall (q,r)  \in  \{1,\dots, Q\}^2$,  $\con_{qr}\in \Acal_\Fcal$,  $\pi_{q} \in (0,1]$ and $\sum_{q  = 1}^Q \pi_{q} = 1$.
The model involves $(Q-1)+Q^2$ parameters, the first term corresponding to the block proportions \new{$(\pi_1 \dots, \pi_{Q-1})$} and the second term to the connection parameters.

However, assuming that the blocks are represented in the same proportions in each network is a strong assumption that may lead to the model being of little practical use. \new{In food webs, the proportion of species at a given trophic level may differ between networks that nevertheless share the same structure, for example networks coming from various studies may have different species resolution (the number of basal species may differ from a network to another).}
%for example  ecosystems may have different number of apex predators.}
The following model relaxes this assumption.

\subsection{A collection of networks with varying block sizes}

$\picolSBM$ still assumes that the networks share a common connectivity structure  encoded in $\balpha$, but that the proportions of the blocks are specific to each network.
More precisely, for $m \in \{1, \dots, M\}$, the $\Xm$ are independent and
\begin{equation}\label{coll:mod:picolsbm}
 \Xm \sim \FSBM_{\nm}(Q, \bpim,\balpha) \quad \quad  \quad \forall m=1, \dots M.
\eqname{$\picolSBM$}
\end{equation}
where $\forall (q,r)  \in  \{1,\dots, Q\}^2$,  $\alpha_{qr} \in \Acal_\Fcal$ and $\sum_{q =1}^Q \pim_{q} = 1,  \forall m \in \{1, \dots, M\}$.
\new{In order to make the model more flexible and suitable    to ecological networks, we allow some block proportions  $\pim_q$  to be null in certain networks ($ \pim_q \in [0,1]$): if $\pim_q=0$ then  block $q$ is not represented in network $m$. The connectivity structure of each network is then part of a larger connectivity structure common to all networks of the collection.}

In order to ensure the identifiability of the model, we assume that each block $q$ is represented in  at least one network $m$, or equivalently,   for any block $q\in\{1, \dots,Q\}$, $\exists m \in \{1,\dots,M\}$ such that $\pim_ q >0$. Let $S$ be the   $M \times Q$ support matrix such that $\forall (m,q)$
\begin{equation*}\label{coll:eq:support}
S_{mq} = \mathbf{1}_{\pim_q>0}.
\end{equation*}
Then, the set of admissible supports is the set of binary matrices with at least one $1$ in each row and column:
\begin{equation*}\label{coll:eq:supportSet}
 \mathcal{S}_Q := \left\{S \in \mathcal{M}_{M, Q}(\{0,1\}),  \sum_{m=1}^M S_{mq} \geq 1   \quad  \forall q = 1, \dots, Q, \sum_{q=1}^Q S_{mq} \geq 1   \quad  \forall m = 1, \dots, M\right\}
\end{equation*}
%Moreover, we assume that  $\sum_{q =1}^Q \pim_{q} = 1,  \forall m \in \{1, \dots, M\}$ where $\pim_q \in [0,1]$  and if $\pim_q=0$ then the  block $q$ is not represented in network $m$. In addition, we assume that for any $q\in\{1, \dots,Q\}$, $\exists m \in \{1,\dots,M\}$ such that $\pim_ q >0$, meaning that any block $q$ is represented  in at least one network.
%\noindent Let $\Qcal_m$ be the indices of the non-null $(\pim_q)_{q=1,\dots,Q}$:
%\begin{equation}\label{coll:eq:Qcalm}
%\Qcal_m = \{q \in \{1,\dots,Q\} | \pim_q>0\}.
%\end{equation}
For a given number of blocks $Q$ and  matrix $S$,  the number of parameters of the $\picolSBM$ model is deduced as follows:
\begin{eqnarray*}\label{coll:eq:Nbpi}
\Nb(\picolSBM) &=& \sum_{m=1}^M \left(\sum_{q=1}^Q S_{mq}  -1\right)  + \sum_{q,r =1}^{Q} \mathbf{1}_{(S'S)_{qr}>0} %\nonumber \\
%&=& \sum_{m=1}^M (|\Qcal_m| - 1)    + \sum_{q,r =1}^{Q} \mathbf{1}_{(S'S)_{qr}>0}.
\end{eqnarray*}
%where $|\Qcal|$ is the cardinal of set $\Qcal$.
The first term  corresponds to the non-null block  proportions in each network. The second quantity
accounts for the fact that some blocks may never be represented simultaneously in any network, so the corresponding connection parameters $\alpha_{qr}$ are not useful for defining the model (see the illustration below).

\begin{remark}
\new{Note that  $\Nb(\picolSBM) \leq M(Q-1)+Q^2$, this upper bound corresponding to the case where all the blocks are represented in all the networks, but with varying proportions.
%Secondly, it is worth to note that if (up to a reordering of the network collection) $S$ has the form
%$$S =  \left(
%\begin{array}{cccc}
% 1_{Q_1}  &   & \dots & 0 \\
%   & 1_{Q_2} &   &  \\
%  \vdots   &  &   \ddots          &        \\
%  0  &   & \dots & 1_{Q_M}
%\end{array}
%\right) $$
%}
%then the $\picolSBM$   is equivalent to the $\sepSBM$.
 }
\end{remark}

\paragraph*{Illustration}
We illustrate the flexibility of $\picolSBM$ model with three examples, all with $Q = 3$ and $M = 2$.
\begin{enumerate}
  \item First consider the situation where the $3$ blocks are represented in the two networks but with different block proportions:
  \begin{eqnarray*}
    \con = \begin{pmatrix}
      \alpha_{11} & \alpha_{12} & \alpha_{13} \\
      \alpha_{21} & \alpha_{22} & \alpha_{23} \\
      \alpha_{31} & \alpha_{32} & \alpha_{33}
    \end{pmatrix} & &  \begin{aligned}
                      \pi^{1} = [.25, .25, .50] \\ \pi^{2} = [.20, .50, .30]
                    \end{aligned}\, .
  \end{eqnarray*}
In that case, $S = \left(\begin{array}{ccc} 1&1&1\\1&1&1 \end{array} \right)$ and the number of parameters is $2(3-1) + 3\times3$ = 13.

    \item \new{Now imagine two networks with nested structures. Blocks $1$ and $3$  are represented in the two networks while block 2 only exists in network $1$. In this illustration, block $2$ may refer to a block of parasites which are not always included in food webs \citep{lafferty2008parasites}}. %Nodes of the first network may populate the $Q = 3$ blocks, while the second network can not populate the second block:  %The number of parameters is given by $(Q-1) + (Q-2) + Q(Q+1)/2$:
      \begin{eqnarray*}
    \con = \begin{pmatrix}
      \alpha_{11} & \alpha_{12} & \alpha_{13} \\
      \alpha_{21} & \alpha_{22} & \alpha_{23} \\
      \alpha_{31} & \alpha_{32} & \alpha_{33}
    \end{pmatrix} & &  \begin{aligned}
                      \pi^{1} = [.25, .25, .50] \\ \pi^{2} = [.40, \text{ }0\text{ }, .60]
                    \end{aligned}\, .
  \end{eqnarray*}
In that case, $S = \left(\begin{array}{ccc} 1&1&1\\1&0&1 \end{array} \right)$ and the number of parameters is $(3-1) + (2-1)+ 3\times3$ = 12.

\item \new{Finally, let us consider two networks with partially overlapping structures. The two networks share block $1$ (for instance basal species)  but the remaining nodes of each network cannot be considered as equivalent in terms of connectivity. One may think of species belonging to trophic chains with different connectivity patterns.} %In other words, nodes of the first network may populate the $2$ of the $Q = 3$ blocks, while the second network is also represented in $2$ blocks but the $2$ networks only share $1$ block.
\begin{eqnarray*}
    \con = \begin{pmatrix}
      \alpha_{11} & \alpha_{12} & \alpha_{13} \\
      \alpha_{21} & \alpha_{22} & \cdot \\
      \alpha_{31} & \cdot & \alpha_{33}
    \end{pmatrix} & &  \begin{aligned}
                      \pi^{1} = [.25, .75, \text{ }0\text{ }] \\ \pi^{2} = [.40, \text{ }0\text{ }, .60]
                    \end{aligned}\, .
  \end{eqnarray*}
In that case,  $S = \left(\begin{array}{ccc} 1&1&0\\1&0&1 \end{array} \right)$. Moreover,  blocks $2$ and $3$ never interact because their elements do not belong to the same network and so $\alpha_{23}$ and $\alpha_{32}$ are not required to define the model. As a consequence, the number of parameters is equal to $(2-1) + (2-1) + 7 = 9$.
\end{enumerate}

\subsection{A collection of networks with varying density ($\denscolSBM$)}
% \PB{}{je ne sais pas si je commencerai tout de suite par l'argument du sampling, pcq il peut y avoir des réseaux plus dense que d'autres sans que ce soit une conséquence de l'échantillonnage, peut être à dire à la fin de ce paragraphe this may be due to different sampling effort}

The $ \iidcolSBM$ can be relaxed in another direction, assuming that the $M$ networks
exhibit similar intra- and inter- blocks connectivity patterns but with different densities. More precisely, let $\dens_m \in \mathbb{R}$ be a  density parameter for network $m$.  The $\denscolSBM$ is defined as follows: %. The $\Xm$ are independent and $\forall m \in \Mcal$
\begin{equation}\label{coll:mod:deltacol}
 \Xm \sim \FSBM_{n_m}(Q, \bpi,\deltam   \balpha).
 \eqname{$\denscolSBM$}
\end{equation}
with
%for all $m \in \Mcal, \quad  1 \leq i < j \leq n_m$, $(q, r) \in \mathcal{Q}^2$:
%\begin{eqnarray*}
%  & \pr(\Zm_{iq} = 1) = \pi_{q},&  \pi_{q} > 0 \text { and } \sum_{q \in \mathcal{Q}} \pi_{q} = 1  \\
%  & \pr(\Xm_{ij}  = 1 | \Zm_{iq}\Zm_{jr} = 1) = \dens_{m}\con_{qr} \in (0,1 ),&\qquad \nonumber
%\end{eqnarray*}
$\pi_q>0, \forall q= 1, \dots,Q$,  $\sum_{q =1}^Q \pi_{q} = 1$. Moreover  $\forall (m,q,r)$, $\deltam \con_{qr} \in \Acal_\Fcal$  and one of the density parameter equal to one ($\dens_1 = 1$) for  identifiability purpose. This model mimics different intensities of connection between networks. \new{In ecology,  these differences in densities between networks could  be due to different sampling efforts for instance, leading to varying total numbers of observed interactions}.
$\denscolSBM$ involves
$\Nb(\denscolSBM) = (Q - 1)+  Q^2 + (M-1)  $ parameters.

\subsection{Collection of networks with varying block sizes and density ($\denspicolSBM$)} Finally, we propose to mix the   models  $\picolSBM$ and $ \denscolSBM$  to obtain a more complex one which allows each network to have its own block proportions $\bpim$ as well as a specific scale density parameter $\deltam$.  Then,  the $(\Xm)_{m \in \{1, \dots, M\}}$ are independent and
\begin{equation}\label{coll:mod:delta-pi}
 \Xm \sim \FSBM_{\nm}(Q, \bpim,\deltam   \balpha),
 \eqname{$\denspicolSBM$}
\end{equation}
where $\forall (m,q,r)$, $\deltam\con_{qr} \in \Acal_\Fcal$, $\delta_1=1$,  $\pim_q \geq 0$  and  $\sum_{q=1}^Q \pim_{q} = 1$. The number of parameters is given by
\begin{equation*}%\label{coll:eq:Nbdeltapi}
\Nb(\denspicolSBM) = \Nb(\picolSBM)+ M-1,
\end{equation*}
the last term corresponding to the aditional proper density of each network. Note that  $\Nb(\denspicolSBM) \leq M(Q-1) +   Q^2 +  M - 1  = MQ +   Q^2 - 1. $  % , where  $\nu(S) = \sum_{q \leq r =1}^{Q} \mathbf{1}_{\exists m : \{q, r\} \subset \mathcal{Q}_m}$ is the number of parameters in $\con$.

%When talking about several $\colSBM$s, we will note the possible models in parenthesis, for instance $(\pi\text{ -} \dens ) \colSBM$s stands for
%$\picolSBM$ and $\denscolSBM$.

% \subsection{Summary} % and illustration}
%
%

% Table \ref{coll:tab:summarymodels} summarizes the five models defined above.
%
%

\begin{table}
\resizebox{\hsize}{!}{%
\begin{tabular}{|c|l|c|c|}
\hline
\hline
Model name & Block prop. & Connection param. & Nb of param.\\
\hline
$\iidcolSBM$& $\pim_q = \pi_q$, $\pi_q >0$&$\alpham_{qr} = \alpha_{qr}$&$ (Q-1)+ Q^2  $\\
\hline
$\picolSBM$& $\pim_q$, $\pim_q \geq 0$ &$\alpham_{qr} = \alpha_{qr}$& $\leq M(Q-1)+Q^2$ \\
\hline
$\denscolSBM$ &$\pim_q = \pi_q$, $\pi_q >0$&$\alpham_{qr} = \deltam\alpha_{qr}$&$ (Q-1) + Q^2 + (M-1) $\\
\hline
$\denspicolSBM$ & $\pim_q$, $\pim_q \geq 0$  &$\alpham_{qr} = \deltam\alpha_{qr}$&$\leq MQ +   Q^2 - 1 $ \\
\hline
$\sepSBM$ & $\pim_q$, $\pim_q>0$ & $\alpham_{qr}$&$\sum_{m=1}^M (Q_m - 1) +  Q_m ^2$\\
\hline
\end{tabular}}
\caption{\label{coll:tab:summarymodels} Summary of the various models defined in Section \ref{coll:sec:model}. The last line corresponds to modeling separately each network as  presented in Section \ref{coll:sec:sbm}.  }
\end{table}

\section{Likelihood and identifiability of the models}\label{coll:sec:properties}
%\subsection{Properties of the Models}

In this section, we   derive the expression of the likelihood of the most complex  model $\dens\picolSBM$ and provide conditions to ensure the identifiability of the parameters for each of the four models.

\subsection{Log-likelihood expression}
For a given matrix $S$, let $\btheta_S$ be: $$\btheta_S = (\bpi^{1}, \dots \bpi^{M}, \dens_1, \dots,\dens_M, \balpha ) = (\bpi, \bdelta, \balpha),$$
where $ \pi^m_q = 0$ for any $q$ such that $S_{mq}=0$. Let $\Zm_{iq} = \mathbf{1}_{\Zm_i = q}$ be the latent variable such that $\Zm_{iq} = 1$ if node $i$ of network $m$ belongs to block $q$, $\Zm_{iq} = 0$ otherwise. We define $\Zm  = (\Zm_{iq})_{i = 1\,\dots, \nm,q = 1\dots, Q}$. Then the log likelihood is:
\begin{equation}\label{coll:eq:likelihood}
  \ell(\obs;\btheta_S) = \sum_{m=1}^M \log \int_{\Zm}\exp\left \{\ell(\Xm | \Zm; \balpha, \bdelta) + \ell(\Zm;\bpi)\right\}d\Zm,
\end{equation}
where
\begin{equation*}
 \begin{array}{ccl}
  \ell(\Xm | \Zm; \balpha, \bdelta) & = & \sum_{\substack{i,j=1 \\ i\neq j} }^{\nm}\sum_{(q,r)  \in \Qcal_m} \Zm_{iq}\Zm_{jr} \log f\left(\Xm_{ij}; \dens_m\con_{qr}\right) ,\\
  \ell(\Zm; \bpi)& = & \sum_{i=1}^{\nm} \sum_{q\in \Qcal_m} \Zm_{iq} \log \pim_{q}
 \end{array}.
\end{equation*}
with $\Qcal_m = \{q \in \{1,\dots, Q\} | \pi^m_{q} >0\}$ and
$f$ defined as in Equation \eqref{coll:eq:emission_dens}.
The log-likelihood functions of the other models can be deduced from this one, setting $\deltam = 1$  for  $\iidcolSBM$ and $\picolSBM$ and  $\bpi^m = \bpi$ for $\iidcolSBM$ and $\denscolSBM$ with $S$ being a matrix of ones (all blocks are represented in each network).

\subsubsection{Identifiability}

We aim at giving conditions ensuring the identifiability of the models we propose. We aim at proving that if  $\ell(\obs;\btheta) = \ell(\obs;\btheta')$ for any collection $\obs$   then  $\btheta = \btheta'$. The proof  relies on the identifiability for the standard SBM demonstrated by  \citet{celisse2012consistency} and is provided in  Section 1 of the Supplementary Material \citep{supplementary}.
Note that, like any mixture models,  all the models are identifiable up to a label switching of the blocks.

\begin{properties} \label{coll:prop:ident_colsbm}$\;$
    %The models are identifiable under the following asumptions and upto the following conditions:
  \begin{description}

    \item [$\iidcolSBM$] The parameters $ ( \bpi, \balpha)$  are identifiable up to a  label switching of the blocks provided that:
    %\textcolor{red}{A voir les preuves sans $\pi$ si on doit identifier les clusters des différents réseaux}
    \begin{enumerate}[label=(1.{\arabic*})]
      \item $\exists m^* \in \{1, \dots, M\} : n_{m^*} \geq 2Q$,
      %\item $\pi_{q} >0$ for all $q \in  \{1, \dots Q\}$
      \item $(\balpha \cdot \bpi)_{q} \neq (\balpha \cdot \bpi)_{r}$  $\forall (q,r) \in \{1, \dots, Q\}^2, q \neq r $.
    \end{enumerate}

    \vspace{1em}

  \item [$\denscolSBM$] The parameters $(\bpi, \balpha, \delta_1,\dots, \delta_M)$  are identifiable up to a  label switching of the blocks provided that:
  \begin{enumerate}[label=(2.{\arabic*})]
    \item $\exists m^* \in \{1, \dots, M\} : n_{m^*} \geq 2Q$ and $\dens_{m^*} = 1$,
    \item $(\balpha \cdot \bpi)_{q} \neq (\balpha \cdot \bpi)_{r}$  $\forall (q,r) \in \{1, \dots, Q\}^2, q \neq r $.

    \end{enumerate}

    \vspace{1em}

    \item [$\picolSBM$]  Assume that $ \forall m =1, \dots,M, \Xm \sim \Fcal \SBM_{\nm}(Q,\bpi^m,\balpha)$. Let $Q_m = |\Qcal_m| = |  \{q = 1\dots, Q, \pi^m_q >0\}|$ be  the number of non empty blocks in network $m$. Then the parameters  $(\bpi^1, \dots, \bpi^M, \balpha)$ are identifiable up to a  label switching of the blocks under the following conditions:
        \begin{enumerate}[label=(3.{\arabic*})]
        \item $\forall m \in \{1, \dots, M\}  : \nm \geq 2Q_m$,
        \item $\forall m \in \{1, \dots, M\}$, $(\balpha \cdot \bpim)_{q} \neq (\balpha \cdot \bpim)_{r}$  $\forall (q,r) \in \mathcal{Q}_m^2, q \neq r$,
        \item Each diagonal entry of $\balpha$ is unique.
    \end{enumerate}

        \vspace{1em}

     \item [$\denspicolSBM$]  Assume that $ \forall m =1, \dots,M, \Xm \sim \Fcal \SBM_{\nm}(Q,\bpi^m,\delta_m\balpha)$. Let $Q_m = |\Qcal_m| = |  \{q = 1\dots, Q, \pi^m_q >0\}|$ be  the number of non empty blocks in network $m$. Then the parameters  $(\bpi^1, \dots, \bpi^M, \balpha, \delta_1, \dots, \delta_M)$ are identifiable up to a  label switching under the following conditions:

    \begin{enumerate}[label=(4.{\arabic*})]
        \item $\forall m \in \{1, \dots, M\}: \nm  \geq 2|\mathcal{Q}_m|$,
        \item $\dens_1 = 1$,
    \end{enumerate}
    \vspace{1em}

    \noindent If $Q \geq 2$:

    \begin{enumerate}
    \item[(4.3)] $(\balpha \cdot \bpim)_{q} \neq (\balpha \cdot \bpim)_{r}$ for all $(q \neq r) \in \mathcal{Q}_m^2$,
    \item[(4.4)] $\forall m \in \{1, \dots, M\}, Q_m \geq 2 $, %$\forall q \neq r = 1, \dots, Q, \quad \exists m : \{q, r\} \subset \Qcal_m$
    \item[(4.5)] Each diagonal entry of $\balpha$ is unique,
     \end{enumerate}
    \vspace{1em}

    \noindent If $Q \geq 3$:
%         \item There are no triplet $\{q, r, s\} \subset \Qcal$ such that $ \exists c \neq 0 : \con_{qq} = c\con_{rr} = c^2\con_{ss}$
%         \item There exists a permutation of the network indices  $\{1, \dots, M\}$, $\varsigma$ such that
%         \[\forall m \geq 2,  | \Qcal_{\varsigma(m)} \cap \cup_{l : \varsigma(l) < \varsigma(m)} \Qcal_{\varsigma(l)}| \geq 2.\]
%        \item Let $S$ be the graph with $Q$ nodes such that $q \leftrightarrow r$ if $\exists m : \{q, r\} \subset \Qcal_m$ of the two following condition is verified:
 %       \begin{enumerate}
  %        \item There exists a cycle of length $Q$
   %       \item
    %    \end{enumerate}
  \begin{enumerate}
    \item[(4.6)] There is no configuration of four indices $(q,r,s,t)\in\{1,\ldots,Q\}$ such that $\alpha_{qq}/\alpha_{rr}=\alpha_{ss}/\alpha_{tt}$ with $q\neq s$ or $r\neq t$ and with $q\neq r$ or $s\neq t$, % cases with r=t or q=r are disregarded because of assumption 5
\item[(4.7)] $\forall m \geq 2,  | \Qcal_{m} \cap \cup_{l : l < m} \Qcal_{l}| \geq 2.$
    \end{enumerate}
  \end{description}
\end{properties}

\section{Inference of the models}\label{coll:sec:inference}

\subsection{Variational estimation of the parameters}

We now tackle the estimation of the parameters $\btheta_S \in \Theta_S$ for a given support matrix $S$. For ease of reading, the index $S$ is dropped in this section.
The likelihood given in Equation \eqref{coll:eq:likelihood} is not tractable in practice, even for a small collection of networks as it relies on summing over $\sum_{m = 1}^M|{\Qcal_m}|^{\nm}$ terms.
A well-proven approach to handle this problem for the inference of the SBM is to rely on a variational version of the EM (VEM) algorithm. The approach is similar for both Bernoulli and Poisson models.

This is done by maximizing a lower (variational) bound of the log-likelihood of the observed data by approximating $p(\lat|\obs;\btheta)$ with a distribution on $\lat$ named $\Rcal$ issued from a family of factorizable distribution \citep{daudin2008mixture}:
\begin{eqnarray*}\label{coll:eq:vbound}
  \mathcal{J}(\Rcal, \btheta) :=   \E_{\Rcal}[\ell(\obs, \lat ; \btheta)] + \mathcal{H}(\Rcal(\lat)) \leq \ell(\obs;\btheta),
  \end{eqnarray*}
where $\mathcal{H}$ denotes the entropy of a distribution. \new{The variational distribution $\Rcal$ can be fully described by the probabilities  $\tau^{m}_{iq}$  where
\begin{equation}\label{eq:tau}
\tau^{m}_{iq} = \mathbb{P}_{\Rcal}(Z^m_{iq} = 1).
\end{equation} These quantities approximate  the posterior node grouping probabilities.}

The VEM algorithm is a two-step iterative procedure which alternates the variational E-step and the M-step. The E-step
consists in optimizing $\mathcal{J}(\Rcal, \btheta)$ for a current parameter value $\btheta$ with respect to $\Rcal$ constrained to be in the family of factorizable distributions.
And the M-step consists in maximizing $\mathcal{J}(\Rcal, \btheta)$
with respect to $\btheta$ for a given variational distribution $\Rcal$.
%
% \SD{}{Donner les 2 etapes VE: $\widehat{\Rcal}  =\argmax_{\Rcal}  \mathcal{J}(\Rcal, \widehat{ \btheta})$ and $\widehat{\btheta}  =\argmax _{\btheta} \mathcal{J}(\widehat{\Rcal}, \btheta)$}.
For the $\colSBM$ models, networks can be treated independently during the  E-step, while the M-step serves as a link between the structures of the networks in the collection. For $\denscolSBM$ and $\dens\picolSBM$ when $\Fcal = \mathcal{B}ernoulli$, numerical approximations are needed  as no explicit expression of $\widehat{\bdelta}$ and $\widehat{\balpha}$ can be derived.
Further details of the variational procedure and the expression of the parameters estimators are provided in Section 2 of the Supplementary Material \citep{supplementary}.

\subsection{Model selection}\label{coll:sec:model_selection}

There are two model selection issues.
First, under a fixed $\colSBM$, we aim to choose the number of blocks $Q$ and determine the support matrix of the blocks $S$ for $\picolSBM$ and $\denspicolSBM$. This task is tackled in Subsection \ref{coll:subsec : Q} by introducing a penalized likelihood criterion.
Second, the comparison of the the $\colSBM$ models --each one introducing various degrees of  consensus between the networks-- with  the $\sepSBM$ -- which assumes that each network has its own structure-- is dealt with in Subsection \ref{coll:subsec:consensus choice}.

\subsubsection{Selecting the number of blocks $Q$}\label{coll:subsec : Q}
%The first task is to propose a criterion to select $Q$ for each \colSBM model.

A classical tool to choose the number of blocks in the SBM context is the Integrated Classified Likelihood (ICL) proposed by \cite{biernacki2000assessing, daudin2008mixture}.
ICL  derives from an asymptotic approximation of the marginal complete likelihood $m(\obs,\lat) = \int_{\btheta}\exp\{\ell(\obs,\lat|\btheta)\}p(\btheta)d\btheta$ where the parameters are integrated out against a prior distribution, resulting in  a penalized criterion of the form $  \max_{\btheta} \ell(\obs,\lat;\btheta) - \frac{1}{2} \pen$. In the ICL,  the latent variables $\lat$ are integrated out against an approximation of
$p(\lat | \obs,\btheta)$ obtained via the variational approximation. This leads to the following expression
$$\ICL = \max_{\btheta}\mathbb{E}_{\widehat{\Rcal}}\left[\ell(\obs,\lat;\btheta)\right]- \frac{1}{2} \pen$$
%where the penalty has to be explicited for each \colSBM model.
Using the fact that
$ \mathbb{E}_{\widehat{\Rcal}}\left[\ell(\obs,\lat;\btheta)\right] \approx \ell(\obs; \btheta) - \mathcal{H}(\widehat{\Rcal})$, one understands
that, as emphasized in the literature, ICL favors well separated blocks by penalizing for the entropy of the node grouping.
%\old{clustering}.
However, in this work, our goal is not only to  group the nodes into coherent blocks but also to evaluate the similarity of the connectivity patterns between the different networks. As such we would like to authorize models providing grouping of nodes
%\old{clustering}
that may be more fuzzy by not penalizing for the entropy. This leads to a BIC-like criterion of the form:
\begin{eqnarray*}\label{coll:eq:bic_gen}
  \BICL  &=&  \max_{\btheta} \mathbb{E}_{\widehat{\Rcal}}\left[\ell(\obs,\lat;\btheta)\right] +\mathcal{H}(\widehat{\Rcal}) - \frac{1}{2}\mbox{pen} = \max_{\btheta} \mathcal{J}(\widehat{\Rcal}, \btheta) - \frac{1}{2}\mbox{pen}
\end{eqnarray*}
We now supply the expression of the penalty term for the four models we proposed and discuss possible variations of the criterion.

\subsubsection*{Selection of $Q$ for $\iidcolSBM$ and $\denscolSBM$}

For $\iidcolSBM$ and  $\denscolSBM$, the derivation of the penalty is a straightforward extension of  the classical SBM model, leading to:
\begin{equation*}
  \BICL(\obs,Q)  =       \max_{\btheta} \mathcal{J}(\widehat{\Rcal}, \btheta)- \frac{1}{2}\left[\pen_\pi(Q) + \pen_{\alpha}(Q) + \pen_{\delta}(Q)\right],
\end{equation*}
where
\begin{eqnarray*}
  \pen_\pi(Q)  &=& (Q-1)\log\left(\sum_{m=1}^M n_m\right),\\
 \pen_{\alpha}(Q)  &=& Q^2   \log(N_M),\\
 \pen_{\delta}(Q) &=& \left\{
\begin{array}{ll}
  0 & \mbox{ for $\iidcolSBM$}\\
  (M-1)  \log\left(N_M\right) & \mbox{ for $\denscolSBM$}
  \end{array}
  \right. .
  \end{eqnarray*}
where \begin{equation}\label{eq:Nbinter}
N_M = \sum_{m=1}^{M} n_m(n_m-1)
\end{equation} is the number of possible interactions.
The first term $ \pen_\pi(Q)$ corresponds to the grouping part where the $Q-1$ block proportions have to be estimated from the $\sum_{m=1}^M n_m$ nodes. The terms  $ \pen_{\alpha}(Q)$ and $\pen_{\delta}(Q)$ are linked to the connection parameters.
Finally, $Q$ is chosen as:
$$\widehat{Q} = \argmax_{Q\in \{1,\dots,Q_{\max}\}} \BICL(\obs,Q)\,.$$

\subsubsection*{Selection of $Q$ for $\picolSBM$ and $\denspicolSBM$}

Here, in addition to the choice of $Q$, the collection of support matrices $S$ is considered. %, where $S \in \mathcal{S}_Q$ ($\mathcal{S}_Q$ has been defined in Equation \eqref{coll:eq:supportSet}).
In order to penalize the   complexity of the model space, we introduce a prior distribution on
$S$ defined as follows. Let us introduce $Q_ m = \sum_{q=1}^Q S_{mq}$  the number of blocks represented in network $m$. Assuming independent  uniform prior distributions on the $(Q_m)$'s  and a uniform prior distribution on $S$ for  fixed numbers of blocks  $Q_1, \dots, Q_M$  represented  in each network, we obtain the following prior distribution on $S$:
\begin{eqnarray*}
% p(S) &=& \sum_{Q_1, \dots, Q_M=1} ^Q p(S | Q_1, \dots, Q_M) p(Q_1\dots,Q_M) \\
% &=&   \sum_{Q_1, \dots, Q_M=1} ^Q \frac{1}{ {Q \choose Q_m}} \frac{1}{Q^M}   = \frac{1}{Q^M}  \prod_{m=1}^M  \sum_{Q_m = 1}^Q \frac{1}{ {Q \choose Q_m}} \\
\log p_Q(S)&=& - M \log(Q) - \sum_{m=1}^M \log   {Q \choose Q_m}
\end{eqnarray*}
where ${Q \choose Q_m}$ is the number of choices of  $Q_m$ non-empty blocks among the $Q$ possible blocks in network $m$.
Now, combining   the Laplace asymptotic approximation of the marginal complete likelihood (where the parameters have been integrated out) and introducing the prior distribution  on $S$, we obtain the  following penalized criterion:
{\small
\begin{equation*}\label{coll:eq:icl_gen}
  \eBICL(\obs,Q)  =       \max_S \left [ \max_{\btheta_S \in \Theta_S} \mathcal{J}(\widehat{\Rcal}, \btheta_S)- \frac{\pen_\pi(Q,S) + \pen_{\alpha}(Q,S) + \pen_{\delta}(Q,S)+ \pen_{S}(Q) }{2}\right],
\end{equation*}}
where
\begin{eqnarray*}
  \pen_\pi(Q,S)  = \sum_{m=1}^M\left(Q_m-1\right)\log(n_m), && \pen_{\alpha}(Q,S) = \left(\sum_{q,r=1}^{Q}\mathbf{1}_{(S'S)_{qr} > 0}\right)  \log\left(N_M\right),
\end{eqnarray*}
\begin{eqnarray*}
 \pen_{\delta}(Q,S)  &=& \left\{
\begin{array}{ll}
 0 & \mbox{for $\picolSBM$}\\
(M-1) \log \left(N_M\right) & \mbox{for $\denspicolSBM$}
\end{array}
\right. ,\\
\pen_{S}(Q)&=&- 2 \log p_Q(S),%\left(   M \log Q + \sum_{m=1}^M \log   {Q \choose \sum_{q=1}^Q S_{mq}} \right)
 \end{eqnarray*}
and $N_M$ has been defined in Equation \eqref{eq:Nbinter}.  Finally, $Q$ is chosen as the number of blocks which maximizes the $\BICL$ criterion:%such that:
$$\widehat{Q} = \argmax_{Q\in \{1,\dots,Q_{\max}\}}   \eBICL(\obs,Q).$$
The details about the derivation of this criterion are provided in  Section 3 of the Supplementary Material \citep{supplementary}.

\subsubsection*{Practical model selection}
The practical choice  of $Q$  and the estimation of its parameters are computationally intensive tasks. Indeed,  we should  compare all the possible models  through the chosen model selection criterion. Furthermore, for each model, the variational EM algorithm  should be  initialized  at a large number of initialization points  (due to its sensitivity to the starting point), resulting in an unreasonable computational cost.  Instead, we propose to adopt a  stepwise strategy, resulting in a faster exploration of the model space,  combined with  efficient initializations of the variational EM algorithm.  The procedure we suggest is  given in Algorithm  \ref{coll:algo:icl} and  is implemented in an R-package \textsf{colSBM} available on GitHub: \url{https://github.com/Chabert-Liddell/colSBM}.
To initialize a $\colSBM$ with $Q$ blocks, we first adjust a $\sepSBM$  with $Q$ blocks, then the $Q$ blocks of the $M$ networks must  be associated. This association  step can be done in many ways due to label switching within each network which provides us with a lot of possible initializations. Then, the stepwise procedure explores the possible number of blocks by building on the previously fitted models. Note that when fitting the $\picolSBM$ or the $\denspicolSBM$, the support $S$ has to be determined which is done through an extra-step that consists in thresholding the parameters $\bpim$ related with the block proportions leading to an exploration over the set $\mathcal{S}_Q$.

\begin{algorithm}[!ht]\label{coll:algo:icl}
\KwData{$\obs$ a collection of networks.}
\BlankLine
\Begin(initialization){
-Infer $\sepSBM$  on  $\obs$, with $Q \in [Q_{min}, Q_{max}]$ \\
-Get $\hat{Z}^{m}_{\sepSBM}(Q)$ \\
-Fit $\colSBM$s with VEM starting from  merged $\hat{Z}^{m}_{\sepSBM}(Q)$ (many initializations as a result of permutations within each $\hat{Z}^{m}_{\sepSBM}(Q)$)\\
-Keep the $b$ fitted models with the best $\BICL$ for each $Q$
}
\vspace{0.5em}

\While{$\BICL$ is increasing}{
- Forward loop

\For {$Q=Q_{min}+1,\ldots, Q_{max}$ }{
	- Fit $\colSBM$ with $Q$ blocks from initializations obtained by splitting a block in  models with $Q-1$ blocks\\
	\If{$\picolSBM$ or $\denspicolSBM$}{
	- Fit $\colSBM$ with  $\widehat{S}_{qm} = \mathbf{1}_{\hat{\pi}^m_{q} > t}$ for different value of threshold $t$ \\
	}
	}

-Backward loop\\
\For {$Q=Q_{max}-1,\ldots, Q_{min}$ }{
	- Fit $\colSBM$ with $Q$ blocks from initializations obtained by merging two blocks in  models with $Q+1$ blocks\\
	\If{$\picolSBM$ or $\denspicolSBM$}{
	- Fit $\colSBM$ with  $\widehat{S}_{qm} = \mathbf{1}_{\hat{\pi}^m_{q} > t}$ for different value of threshold $t$ \\
	}
	}

    - Among all fitted models, keep the $b$ fitted models with the highest $\BICL$ for each $Q$\\
   % - Set $Q_{min} - d := \arg\max \eBICL(Q) =: Q_{max} + d$ for a given depth $d$
}
 \Return{$\widehat{Q} = \arg\max \eBICL(\obs,Q)$, with the corresponding  $\widehat{\btheta}$, $\widehat{\mathbf{Z}}$ and $\widehat{S}$ for $\picolSBM$ and $\denspicolSBM$.}
\caption{Model selection algorithm}
\end{algorithm}

%\begin{remark}
%Due to the stochasticity of the V--EM algorithm for $\colSBM$s induces by the random order in the networks for the $E$--step when $M \geq 3$, it may be advisable to relaunch the procedure a few times to avoid unlucky run.
%By default in our implementation, we set the number of run to $3$, the $depth$ of the exploration to $d = 2$, the threshold to $t \in [.01, .05]$ and the number of kept models to $b = 3$.

%\SD{}{J'enlève la remarque sur la stochasticité (cf en commentaire). Par certaine qu'on doive donner ce niveau de détails. par contre donner le package oui!}

%\end{remark}

\subsubsection{Testing common connectivity structure}\label{coll:subsec:consensus choice}

% \PB{}{dire qu'on choisit entre les différents modèles pi delata iid et sbm sans structure commune et que la comparaison naturellement intéressante est par rapport à des sbms sans structures communes}
We can also use a model selection  approach to choose which model from the $4$ $\colSBM$s and the $\sepSBM$ is the most adapted to the collection.
The most interesting comparison is to decide whether a collection of networks share the same connectivity structure by comparing the model selection criterion obtained for a given  $\colSBM$ model with the one of $\sepSBM$. We decide that a collection of  networks share the same connectivity structure if:
\begin{equation*}
  \max_{Q} \eBICL_{\colSBM}(\obs,Q) > \sum_{m =1}^{M}\max_{Q_m} \BICL_{SBM}(\Xm,Q_m).
\end{equation*}

% \begin{remark}
%   For $(\dens\text{-}\dens\pi)\colSBM$s with $Q$ large enough, there always exists $S$ such as $\max_{S} ICL_{\colSBM}(S) = \max_{Q_1} ICL_{SBM}(Q_1) + \max_{Q_2} ICL_{SBM}(Q_2)$, where $|\Qcal_1| = Q_1$, $|\Qcal_2| = Q_2$, $\Qcal_1 \cap \Qcal_2 = \emptyset$. In this case, the $2$ networks have no common blocks and so they do not share any common connectivity structure.
% \end{remark}

\subsection{Simulation studies}
In  Section 4 of the Supplementary Material \citep{supplementary}, we perform a large simulation study in order to test the inference and model selection procedures proposed in this section. More specifically, simulating from a $\picolSBM$ model, for various strengths of connectivity structures, we look at our capacity to recover the true connectivity parameter $\balpha$ as well as grouping of the nodes and the true support $S$. We also assess the quality and the limit of the model selection with the $\BICL$ criterion. First, we test our ability to select the true number of blocks and to distinguish $\picolSBM$ from $\iidcolSBM$ and $\sepSBM$, and second we vary  the block proportions and study its influence on model comparison and the selection of the number of blocks and their support.

We perform another simulation study to understand how, for particular configuration, using a $\colSBM$ model on a collection of networks favors the transfer of information between networks and allows to find finer block structures on the networks. \new{An additional one is dedicated to collection of networks simulated from $\sepSBM$ with heterogeneous numbers of nodes. We study how the different $\colSBM$ models deal with spurious structures and our ability to detect them through the $\BICL$ criterion.}

\section{Partition of a collection of networks according to their  connectivity structures}\label{coll:sec:partition}

If the networks in a collection do not have the same connectivity structure, we aim to cluster them accordingly.
%This is  a central issue in ecology for instance.
\new{In order to do this, we propose to use the $\BICL$ criterion in a similar fashion as we did for testing common connectivity structure in Section \ref{coll:subsec:consensus choice}. We seek the partition of the collection which maximizes a score based on the $\BICL$ criterion. In this partition, each sub-collection of networks has its own structure given by a $\colSBM$, which represents the best way to model the collection according  to the criterion.}
%We present hereafter a strategy to perform a clustebestring of the networks.
%In what follows, we do not specify the type of $\colSBM$ we consider because the strategy is adapted to any of them.

Clustering a collection of networks consists in finding   a partition $\mathcal{G} = (\Mcal_g)_{g  = 1, \dots,G}$ of $\{1, \dots,M\}$. %, such that  $\cup_{g} \mathcal{M}_g = \{1, \dots, M\}$.
Given $\Gcal$, we set the following model on $\obs$:  % and   %with a set of $\Qcal^g$ blocks.

\begin{equation*}
 \forall g \in \{1, \dots, G\}, \quad \forall m \in \Mcal_g, \quad  \Xm \sim \FSBM(Q^g, \bpim, \deltam \balpha^g)
\end{equation*}
with $\dens_1 = 1$. Moreover,   $\dens_{m} = 1 $   for all $ m$ for  $\iidcolSBM$  and $\picolSBM$s
and $\bpim = \bpi^g$ for $\iidcolSBM$ and $\denscolSBM$.
In other words, the networks belonging to the sub-collection $\Mcal_g$ share the same mesoscale structure given by a particular $\colSBM$. % structure.
To any partition $\mathcal{G}$, we associate the following score:
\begin{equation}\label{coll:ICL:partition}
  \Sc(\Gcal) =   \sum_{g  = 1} ^G \max_{Q^g = 1,  \dots, Q_{\max}}\BICL((\Xm)_{m \in \Mcal_g}, Q^g).
\end{equation}
where $\BICL((\Xm)_{m \in \Mcal_g}, Q^g)$ is the $\BICL$ computed on the  sub-collection of networks $\Mcal_g$.
The best partition is chosen as the one which maximizes the score $\Sc(\Gcal)$ in Equation \eqref{coll:ICL:partition}.
%where $  \in \mathcal{P}(A)$ is the set of partitions of $ A$. %\mathcal{P}(\{1, \dots, M\}$
%We expose the method for $\denspicolSBM$.  %Let us fixe a  $\colSBM$ model introducing a consensus structure.
% constraints on the parameters, the method is straightforwardly adapted to the other $\colSBM$ models.

%\subsection{Exhaustive research}

\vspace{1em}

\noindent Computing the $\BICL$ for all the partitions $\Gcal$ requires to consider the  $2^{M}-1$ non-empty sub-collections of the networks $\Mcal$, fit the $\colSBM$s on these sub-collections and then combine the associated $\BICL$ in order to be able to compute the scores given in Equation \eqref{coll:ICL:partition}. %  the models which are involved in a partition of the networks.
This can be done exhaustively provided that $M$ is not too large but the computational cost becomes  prohibitive as $M$ grows.

%The first one relies on the $\BICL$ criteria but can only be applied for a small collection of networks. The second is only adapted to the models with varying blocks size (namely $\picolSBM$ and $\denspicolSBM$). The third strategy is based on the definition of a dissimilarity between networks computed from the estimated parameters.

 To circumvent this point, we propose a less computationally intensive  forward strategy, starting from $\Gcal = \Big\{\{1, \dots,M\}\Big\}$, and then progressively splitting the collection of networks.
In order to explore the space of partitions of $\{1,\dots,M\}$,  we define a dissimilarity measure between any pair of networks $(m,m')$ in a sub-collection. % The dissimilarity is  based on the parameters encoding their mesosclale structure.

\new{This dissimilarity is a squared distance weighted by the block proportions between the connectivity matrices of the two networks. The parameters (block proportions and connectivity matrices) are computed separately on the two networks with the node grouping provided by the inference on the whole sub-collection.
 We then use $2$-medoids clustering to split the sub-collection of networks based on the dissimilarity measures. A split is validated if it increases the score of Equation \eqref{coll:ICL:partition}. The mathematical definition of the dissimilarity measure and details on the recursive clustering algorithm are given in Appendix \ref{coll:ap:sec:partition}.}

% \new{Begin: En appendix ?}
%
% \new{end: In appendix or supplementary}

\paragraph*{Simulation study} We illustrate our capacity to perform a partition of a collection of networks based on their structure for all $4$ $\colSBM$ models in  Section 4 of the Supplementary Material \citep{supplementary}.

\section{Application to Food Webs}\label{coll:sec:foodwebs}

In this section, we   demonstrate the interest of our models on  the two collections of ecological networks described  in Section \ref{coll:sec:sbm}.  %\old{The first one consists of a collection of $3$ stream food webs issued from the dataset of \cite{thompson2003impacts} described in Section \ref{coll:sec:sbm}. We   analyze in detail the different structures given by the different models and we  show how using networks with some common structure helps the prediction of missing information in the networks. The second dataset is a collection of $67$ networks issued from the Mangal database \citep{rmangal}. We will use our model to propose a partition of the collection into sub-collections of networks with common mesoscale structures.}

\subsection{Joint analysis of $3$ stream food webs}
%We select a collection of $3$ stream food webs issued from the work of \cite{thompson2003impacts} in the USA, namely the one from Martins, Cooper and Herlzier with respective number of species $105$, $58$, $71$ and trophic links $343$, $126$, $148$.
%The food webs comprise of directed trophic links showing the energy flow (ie. $X^{m}_{ij} = 1$ if species $j$ prey on species $i$) with no reciprocal interactions.

In Section \ref{coll:sec:sbm}, we fitted $\sepSBM$  and  obtained $5$ blocks for Martins, $3$ blocks for Cooper and $4$ blocks for Herlzier. For reminder, a matricial representation of the block reordered food webs was shown in Figure \ref{coll:fig:3net_sbm}. Each food web has $2$ blocks of basal species (the $2$ bottom blocks). %For Cooper, the higher trophic level are all in the same block, but lack the statistical power to further refine the species clustering.  For Herlzier the higher trophic level is separated into $2$ blocks mainly determined on how much they prey on the less preyed basal block. Martins has a separation into $3$ blocks, the $3$rd one is a medium trophic level, which prey on basal species and are highly preyed by species of the $1$st block. The first two blocks comprise of higher trophic species with the $2$nd blocks being much less connected than the $1$st one.

\paragraph*{Finding a common structure between the networks}

We now fit the four $\colSBM$ models  in order  to find a common structure among the $3$ networks. First,  notice that our model selection criterion greatly favors common network structure above separated one: $\BICL = -2080$ for $\sepSBM$ versus respectively $-1964$, $-1983$, $-1970$ and  $-1988$ for $\iidcolSBM$, $\picolSBM$, $\denscolSBM$, $\denspicolSBM$.
The \new{mesoscale} structures of the collection under the different models are represented  in Figure \ref{coll:fig:3net_colsbm}. In this figure, \new{the red shaded matrices represent the estimated connectivity matrices $\widehat{\balpha}$. In these matrices, the sizes of the blocks are proportional to the block proportion parameters $\widehat{\bpi}$ for $\iidcolSBM$ and $\denscolSBM$ and proportional to
the averages of block proportion parameters over the networks $\widehat{\bpim}$ weighted by the number of species within the networks.} %computed as the average of each network block proportion}. % \sout{the square matrix represents the estimated connection matrix $\widehat{\balpha}$}
The cumulative bar plot on the right represents \new{the actual block proportions for each network resulting from the inference of a $\colSBM$ denoted as $\widetilde{\bpi}^m$ where $$\widetilde{\pi}_q^m = \sum_{i=1}^{n_m}  \widehat{\tau}_{iq}^m.$$ Note that although the $\iidcolSBM$ and $\denscolSBM$ assume common parameters for block proportions, the actual block proportions of each network  $\widetilde{\bpi}^m$ fluctuate around the block proportions $\widehat{\bpi}$ .}% \pageref{coll:page:tilde})}.

%\textcolor{red}{Compliqué de mettre dans un graphe ce qui n'a pas été défini? (mis en annexe finalement. }

%\textcolor{red}{ne pas mettre de référence aux numéro de page. ca n'est pas possible dans un journal}

For each model, the basal species are separated into $2$ blocks (bottom blue blocks in Figure \ref{coll:fig:3net_colsbm}), similar to the one obtained with the SBMs. \new{For higher trophic levels, the inferred structures slightly differ and are more detailed for the Cooper and Herlzier networks than the ones obtained with SBMs as described hereafter.}%\sout{The obtained structure slightly differs as described hereafter} %\textcolor{red}{Pourquoi avoitr change cette phrase?}\textcolor{blue}{Pour mettre en avant la comparaison avec les resultats du SBM, cf reviewer 1}.
\begin{itemize}
  \item $\iidcolSBM$ highlights  $5$ blocks in total. Block $3$ (light green) is a small block of intermediate trophic level species (ones that prey on basal species and are being preyed on by higher trophic levels) with some within block predation. The higher trophic level is divided into $2$ more blocks,  block $2$ (dark green) only preys on the 2 basal blocks, while block $1$ (pink) preys on the intermediate block $3$ level but only on the most connected basal species block.
  \item $\picolSBM$ leads also to  $5$ blocks.%, with blocks $1$ and $2$ corresponding to the top and intermediate trophic levels.
  There are no empty blocks and the block proportions are roughly corresponding to the ones of $\iidcolSBM$. This renders the flexibility of the $\picolSBM$ of little use compared to the $\iidcolSBM$ on this collection.
  \item With $\denscolSBM$, the species are grouped into $6$ blocks and the networks have different estimated density parameters: $\hat{\delta} \approx (1, .9, 1.2)$. Block $1$ (red), $2$ (pink) and $3$ (dark green) correspond to the top trophic levels. Block $4$ (light green) is an intermediate trophic level group, well connected with both block $1$ and the basal species blocks. Block $2$ (pink) is huge and only preys on block $6$ while block $3$ (dark green) is a small group of species that preys on both basal species blocks.
  \item Finally, $\denspicolSBM$ groups the species into $5$ blocks %with more heterogeneous block proportions
  and the networks have different estimated density parameters: $\hat{\delta} \approx (1, .9, 1.4)$. The connectivity structure is almost similar to the one of $\picolSBM$ but the blocks have different proportions. Block $1$ (pink) corresponds to block $1$ (red) and $2$ (pink) of $\denscolSBM$ and is the merge of two top trophic levels. Again, the block proportions are still quite homogeneous between networks, and the added flexibility of $\denspicolSBM$ compared to $\denscolSBM$ is not used.
\end{itemize}

\begin{remark}
  On this collection, the entropy of the block memberships is much lower on $\denspicolSBM$ than on $\denscolSBM$. In this model,  to ensure  homogeneous block proportions between networks, some nodes tend to get a fuzzy grouping and sit between several blocks (the variational parameters do not concentrate on one block). This phenomenon is taken into account by our model selection criterion which tends to favor models with higher entropy than models with well separated blocks.
\end{remark}

% \pagebreak

{\centering{
\begin{figure}[!t]
    \includegraphics[width=\hsize]{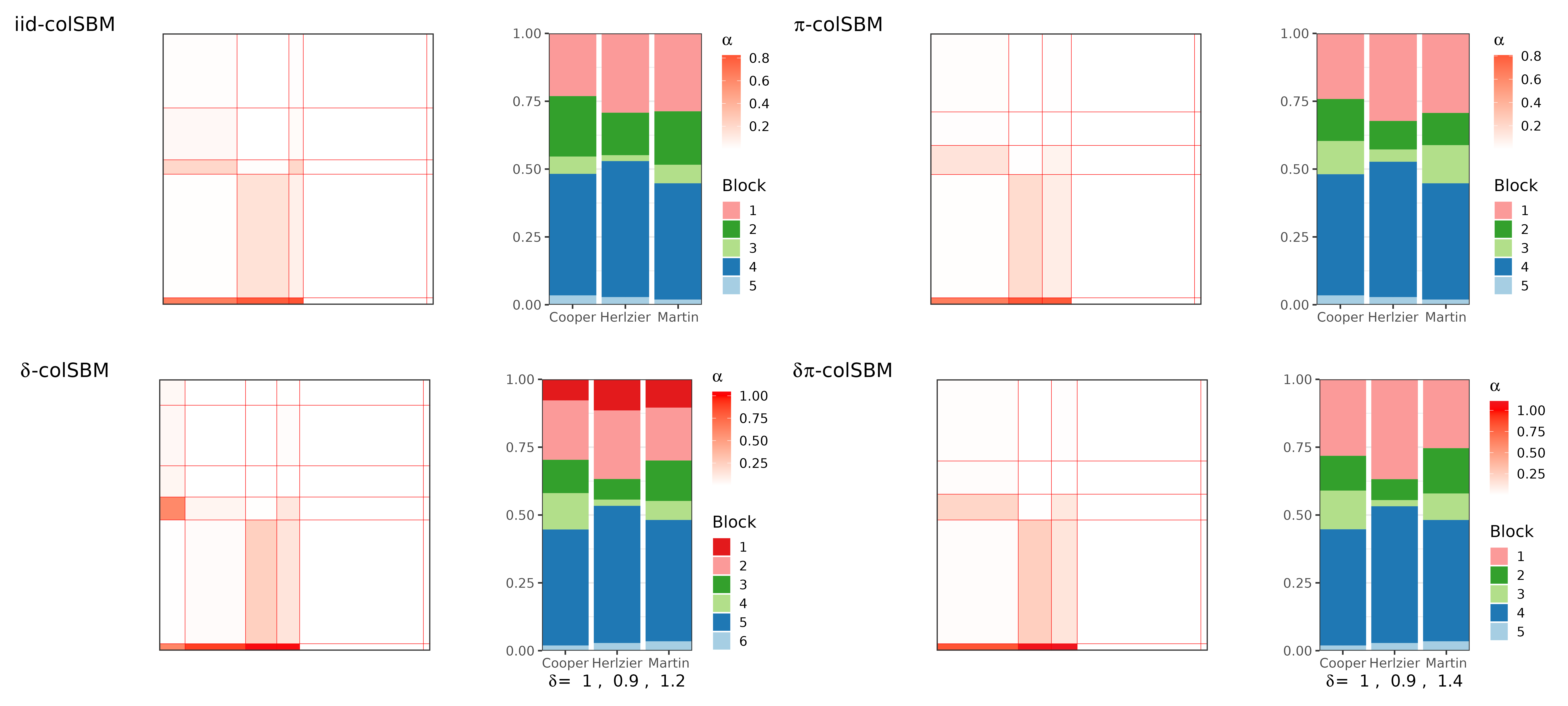}
  \caption{\textbf{Estimated structure of the collection of $3$ stream food webs with the four $\colSBM$ models}.
  For each model, the matrix on the left is the estimated connectivity parameter  $\widehat{\balpha}$. \new{The sizes of the blocks are proportional to the block proportion parameters $\widehat{\bpi}$ for $\iidcolSBM$ and $\denscolSBM$ and proportional to
the averages of block proportion parameters over the networks $\widehat{\bpim}$ weighted by the number of species within the networks.}. The barplot on the right depicts $\widetilde{\bpi}^{(m)}$. The ordering is done by trophic level from bottom to top and right to left. For $(\dens \text{-} \dens\pi)\colSBM$s we give $\hat{\dens}$ below the barplot.}\label{coll:fig:3net_colsbm}
\end{figure}
}}

\paragraph*{Prediction of missing links and dyads}
Since we have been able to  find some common structures between the $3$ networks, we now examine if these structures could be used to help recover some information on networks with incomplete information.
We proceed as follows: we choose a network $m$ and remove a proportion $K \in [.1, .8]$ of
\begin{itemize}
 \item the existing links uniformly at random  \emph{for the missing link experiment}
 \item or of the existing dyads (both $0$ and $1$) by encoding them as \texttt{NA} \emph{for the missing dyads experiment}.
\end{itemize}
Then, for  the missing link experiment, we   try to recover  where  the missing links are among all non existing ones. For the missing dyad experiment, we predict  the probability of existence of missing dyads (\texttt{NA} entries).
Under the $\colSBM$, the probability of a link between species $i$ and $j$ for network $m$ is predicted by:
\begin{equation*}
  \widehat{p}^{m}_{ij} = \sum_{q,r \in \hat{\Qcal}_m} \widehat{\tau}^{m}_{iq}\widehat{\tau}^{m}_{jr}\widehat{\dens}^{m}\widehat{\con}_{qr},
\end{equation*}
where $\tau^m_{iq}$ and $\tau^m_{jr}$ are defined as in Section \ref{coll:sec:inference}.
We resort to the area under the ROC curve to evaluate the capacity of the different models to recover this information. %As the networks are very sparse, we will use the Poisson model for $(\dens\text{-}\dens\pi)\colSBM)s$. It provides a good approximation of the Bernoulli's one, and we have an explicit expression for the M-step of the variational algorithm.
For each value of $K$, each experiment is repeated $30$ times and the results are shown in Figure \ref{coll:fig:predict}.
{\centering{
\begin{figure}[!t]
    \includegraphics[width = \hsize]{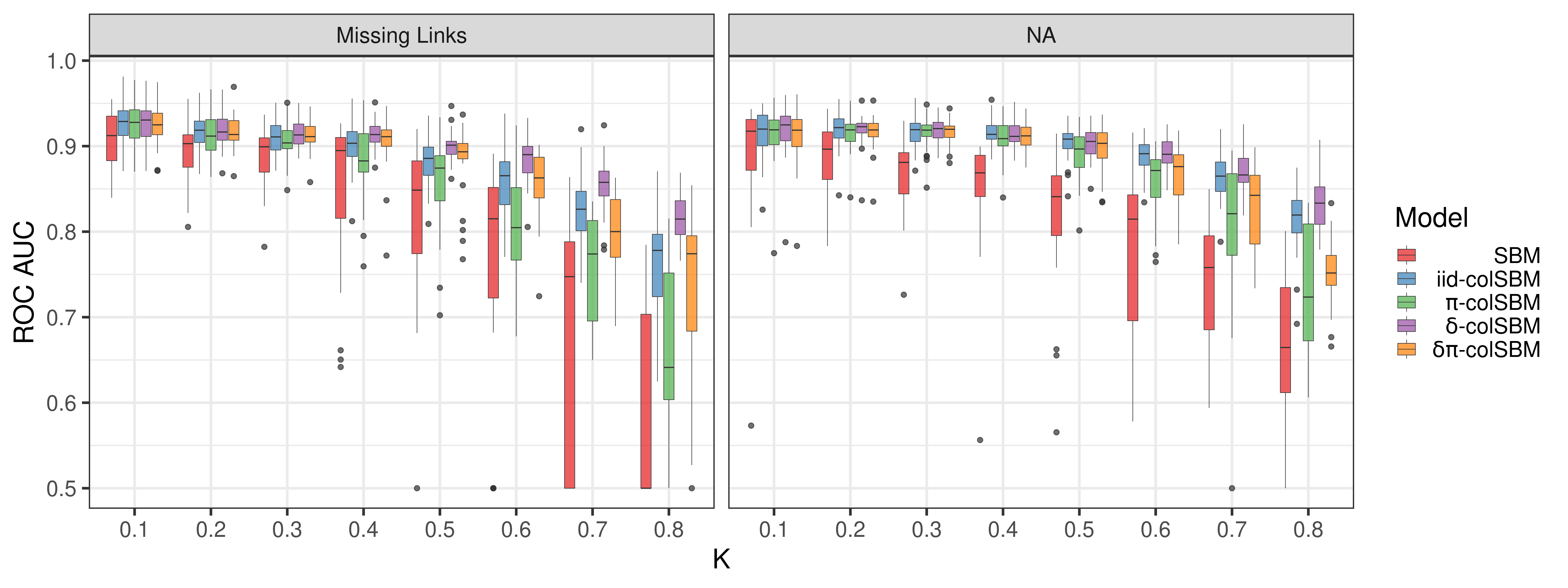}
  \caption{Prediction of missing links and \texttt{NA} entries on stream food webs.}\label{coll:fig:predict}
\end{figure}
}}

First, let us notice that these stream food webs networks have a structure that is well explained by an SBM. When there is little information missing ($K < .3$) the ROC AUC is over $0.9$. Besides, with 70$\%$ of missing links or dyads,  we still predict better than a random guess (ROC AUC  $\approx .75$). As there is a common structure between the $3$ networks, there is a lot of information to be taken from the ones with no missing information.  Starting from $K \geq .2$, the $\colSBM$s outperform the $\sepSBM$ on both experiments. Even for $K = .8$, the prediction is still high. About  the difference between the $\colSBM$s, for the missing links experiment, as we remove links from one of the network, its density decreases and the models with a density parameters, $\denscolSBM$ and $\denspicolSBM$  have a built-in mechanism that compensates this fact. As a consequence,  these models yield to better predictions than  $\iidcolSBM$ or $\picolSBM$ for large values of $K$. %This phenomenon is less clear on the \texttt{NA} experiment.

Another noteworthy comment on both the \texttt{NA} and missing links experiments is that as $K$ grows, the amount of information on the modified network gets lower. Hence, $\picolSBM$ and $\denspicolSBM$ lack the statistical power to separate blocks and will empty blocks on this network. This affects our capability to predict the trophic links. On the other hand, $\iidcolSBM$ and $\denscolSBM$ will force some separation of the species into blocks, and as the information from the other networks is relevant it still has good prediction performance for large $K$.

\subsection{Partition of a collection of $67$ predation networks}
Now, \new{we consider the collection of networks issued from the Mangal database \citep{rmangal} introduced at page \pageref{coll:sepSBM:predation} of Section \ref{coll:sec:sbm}.}
%\old{a collection of $67$ predation networks which are all directed networks with more than $30$ species, from the Mangal database \citep{rmangal}. They are issued from $33$ datasets each containing between  $1$ and $10$ networks. The number of species ranges from $31$ to $106$ ($3395$ in total) by networks; the networks   have density ranging from $.01$ to $.32$ ($14934$ total predation links).}
This dataset is too heterogeneous to find a common structure that will fit well on all the networks. Therefore we propose to use a $\picolSBM$ to look for a partition of the networks into groups sharing  common connectivity  structures.
We focus our investigations on this model since we aim to cluster together networks which share some blocks with similar features but we do not expect all the networks in a sub-collection to share exactly the same blocks.
%A supplementary  application on the same dataset using a $\denscolSBM$  is given in Section 5 of the Supplementary Material \citep{supplementary}.

\new{Fitting a $\picolSBM$ on the whole collection provides a $13$ blocks connectivity structure with more than half of the blocks being empty  ($457/871$). It leads to a much higher $\BICL$ than the one given by $\sepSBM$ ($-31303$ vs. $-33311$), still the partition we provide below greatly improves this criterion ($-30703$).
%We show the connectivity structure and the blocks support in Figure \ref{coll:fig:full_rmangal}.
In this partition, the networks are clustered  into $5$ sub-collections. The obtained partition and the connectivity structure of each sub-collection are shown in Figure \ref{coll:fig:classif_rmangal} as well as a contingency table of the obtained sub-collections crossed with the different datasets of the Mangal database. The number of blocks of each sub-collection varies between $8$ and $12$. The sub-collection  denoted \textbf{A} in the following has $12$ blocks and contains networks that are denser (networks density ranging from $0.17$ to $0.32$) than the networks of the other sub-collections (networks density ranging from $0.01$ to $0.17$). Each sub-collection contains between $1$ and $3$ blocks which can be considered as blocks containing mainly basal species. Indeed, those blocks have a very low in-going  interaction probability with all the other blocks (inferior to $0.02$). All the structures exhibit low within blocks connectivity for most blocks ($\widehat{\alpha}^{g}_{qq}$ is small for most $q$), meaning that predation links between species of the same block are unlikely, with the exception of the sub-collection \textbf{A}.  Also, the networks in all the sub-collections contain mostly trophic chains and only a few cycles, since each of the connectivity parameter matrices $\widehat{\balpha}^{g}$ can almost be reordered as  triangular matrices.
Our detailed comments on each sub-collection follow.}

%
% \begin{figure}[!t]
% {\centering{
%     \includegraphics[width=.8\hsize]{collection/rmangal_full_graphon.png}
%   \caption{Left: Support $S$ of the blocks (white for absent) provided by $\picolSBM$ for the $67$ predation networks of the collection issued from the Mangal database. The networks are reordered from bottom to top according to the partition of the collection.  Right: Mesoscale structure of the same collection.}\label{coll:fig:full_rmangal}
% }}
%   \end{figure}

%{\color{olive}
\begin{itemize}
  \item [\textbf{A}] This sub-collection consists of $7$ networks and $12$ blocks are required to describe this sub-collection. $5$ networks are issued from the same dataset (id: 80). These $5$ networks populate the $12$ blocks, while the other $2$ networks only populate parts of them. The average density is about $0.18$. From the ecological point of view, the blocks can be divided into $3$ heterogeneous sets: block $1$ to $3$ represent the higher trophic levels, block $4$ to $8$ the intermediate ones and block $9$ to $12$ the lower ones.

  \item [\textbf{B}] A sub-collection of $26$ networks with heterogeneous size and density issued from various datasets. Most networks populate only parts of the $8$ blocks. The structure is mainly guided by blocks $2$, $3$, $5$ and $6$, from higher to lower trophic levels.  Block $4$ is represented in only $5$ networks where it is either an intermediate or a bottom trophic level.  It introduces some symmetry in the connectivity matrix rendering it difficult to order the blocks by trophic order. Species from top trophic levels prey on basal species.

  \item [\textbf{C}] A small sub-collection of $6$ networks with density ranging from $.06$ to $.11$. All networks are represented in $5$ or $6$ of the $7$ blocks, including the first three blocks. The sub-collection consists of $3$ of the $5$ networks of dataset $48$, the separation being based on the collecting sites. The top trophic level is divided into $2$ blocks, species from those blocks preying only on intermediate trophic level species. One can exhibit two different trophic chains: species from block $2$ prey on  species from block $4$, which prey more on basal species (block $7$) than on others intermediate trophic species (block $6$), while species from block $1$ prey on species from block $3$ and $4$, block $3$ exhibiting the inverse behavior of block $4$.

  \item [\textbf{D}] Another heterogeneous sub-collection of $23$ networks. The $10$ networks from dataset $157$ (stream food webs from New Zealand) are divided between sub-collections \textbf{B} and \textbf{D} based on the type of ecosystem. The data from sub-collection \textbf{B} were collected in creeks, while the one from sub-collection \textbf{D} were collected on streams. Compared to the other heterogeneous sub-collection (\textbf{B}), the top trophic species (block $1$, $2$ and $3$) prey mostly on intermediate trophic levels (block $4$ and $5$). Species from block $4$ prey on species from the other intermediate block and the $3$ blocks of bottom trophic level species (blocks $6$, $7$ and $8$), while species from block $5$ just prey on species from the last two blocks.

  \item [\textbf{E}] The last sub-collection consists of $6$ networks from various datasets. In the $7$ blocks structure, the species of block $1$ (represented on $4$ of the $6$ networks) prey on species from all other blocks with the exception of block $7$. The basal species are separated between blocks $6$ and $7$ depending on whether or not they are preyed on by species from the first two blocks.
\end{itemize}%}

{\centering{
\begin{figure}[!t]
    \includegraphics[width=\hsize]{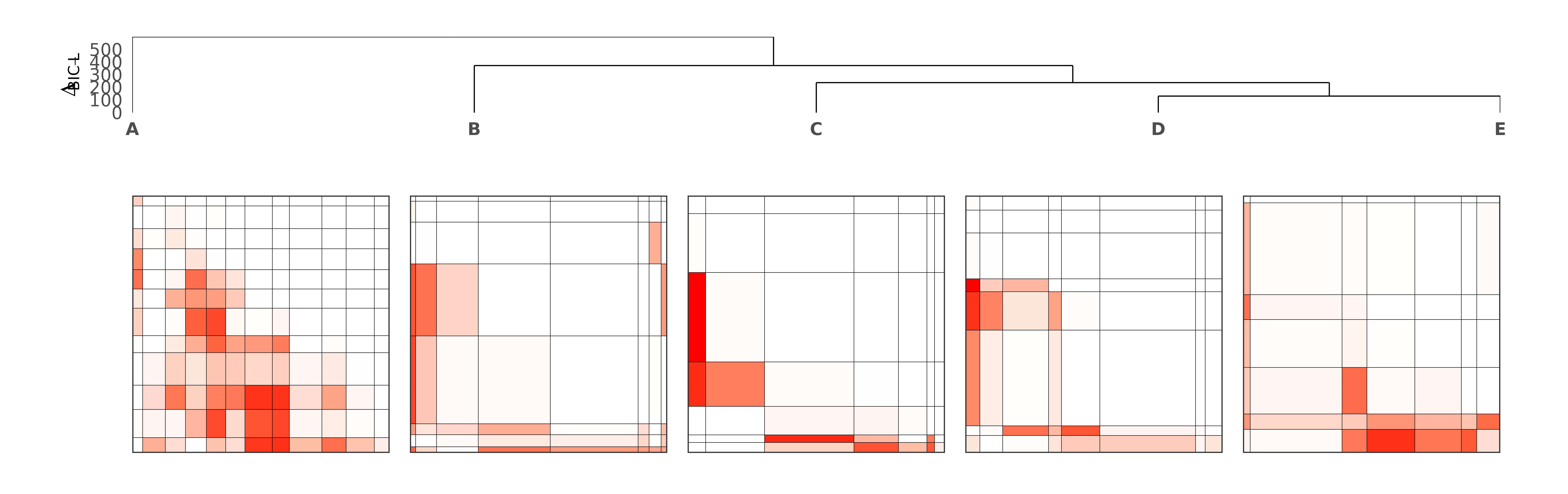}
        \centering{\includegraphics[width=.8\hsize]{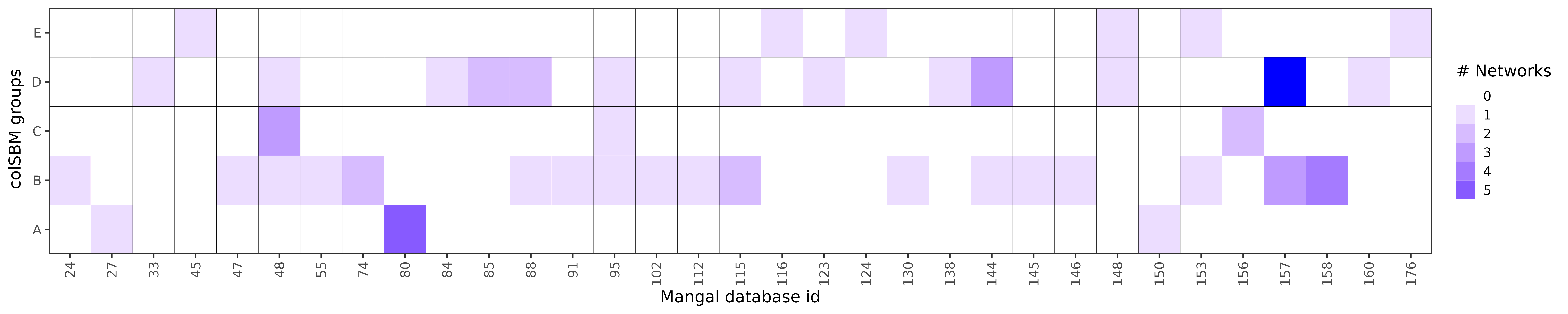}}
  \caption{Above: Clustering and connectivity structures of a collection of $67$ predation networks from the Mangal database into $5$ sub-collections. The length of the dendrogram is given by the difference in $\BICL$ to the best model. Below: Contingency table of the clustering found by $\picolSBM$ and the different datasets from the Mangal database.}\label{coll:fig:classif_rmangal}
\end{figure}
}}

\section{Discussion}
In this paper, we proposed a new method to find a common structure and compare different structures of a collection of networks which we do not assume to share common nodes. This method is very general and could be applied to networks sharing common nodes as well, such as temporal or multiplex networks. Starting from our most basic model, -- namely the $\iidcolSBM$ -- we refined it by proposing models allowing for different mixture distributions and even empty blocks ($\picolSBM$), models allowing to find common structure for networks of different density ($\denscolSBM$) or even models allowing both ($\denspicolSBM$).
\new{These are only a few of the possible models within the $\colSBM$ framework. Other systems of constraints on the SBM parameters can be proposed.
For instance, when driven by the analysis of common community structures, one may impose the diagonal of the connectivity parameter matrix (which corresponds to the within block connectivities) to be common for all the networks but
release the restriction on the off-diagonal parameters (between blocks connectivities).}
The model selection criterion we derived can be used to select the number of blocks but  also to choose which $\colSBM$  fits better  the data.  We also presented a strategy providing a  partition  of a collection of networks into sub-collections of networks sharing  common connectivity patterns. % given by the adapted $\colSBM$.

% An application that we leave for future work, but which motivated our modeling choice is the ability to compare the nodes clustering of different networks obtained under a common ``rule". That is instead of looking at the global patterns as we did in this paper, this method could be used to retrieve which species/individuals play the same role on different networks, in this case our ability to determine if using a $\colSBM$ over a collection of $SBM$s is meaningful is of particular importance. This aspect could be further refined by proposing a test for a fixed number of blocks, where the null hypothesis would be that the networks share common structures. If the null hypothesis is rejected, then the blocks obtained with a $\colSBM$ should not be analyzed any further.

The idea behind these models is very general and could be extended to other types of networks. In ecology, bipartite and multipartite networks are common and the model extension \citep{govaert2003clustering, bar2020block} is straightforward (although some additional modeling choices  arise when considering $\picolSBM$, $\denscolSBM$ or $\denspicolSBM$), the main difficulty would then  lie   in the algorithmic part.
\new{Indeed, the dimension of the model space is larger when considering blockmodels for multipartite networks: ${\mathbb{N}^{*}}^{K}$ instead of $\mathbb{N}^{*}$ as a number of blocks has to be determined for each of the $K$ different types of  units. Thus, our model selection algorithm will not scale well.
An adaptation of \citet{come2021hierarchical} to the colSBM with the derivation of an exact ICL criterion and a genetic algorithm to search in the model space  could be a direction to solve this issue.}
% \new{as our model selection algorithm does not scale well with the dimension of the model space (${\mathbb{N}^{*}}^{K}$ instead of $\mathbb{N}^{*}$, where $K$ is the number of different types of units in a $K$-partite network). Adapting the exact ICL with a genetic algorithm  to the $\colSBM$ extensions might be a way to solve this problem.}.
Additionally, incorporating the type of ecological interaction as a network covariate \citep{mariadassou2010uncovering} would help us understand its impact on the structure of the networks and to the robustness of the ecosystems they depict \citep{chabert-liddell2022robustness}.  The main idea of this article could also be extended to the Degree Corrected $SBM$ \citep{karrer2011stochastic} which is quite used in practice. %Finally, the nested version of the SBM proposed by \cite{peixoto2014hierarchical}, by allowing a hierarchy on the blocks would be particularly adapted to $\picolSBM$ and $\denspicolSBM$s.

Finally, we notice during our simulations and applications that the $\colSBM$s allow  to find a larger number of blocks compared to the $\sepSBM$ and so lead to a finer resolution of the mesoscale structure of the networks. This resolution limit problem was one of the motivations of \cite{peixoto2014hierarchical} and we believe that this direction should be explored further for collections of networks. \new{Also, sometimes practitioners are not interested in the model selection procedures but seek the structure and nodes grouping for a set number of blocks. In the case where networks strongly vary in sizes or reliability of information, it might be interesting to use a weighted likelihood as  \citet{vskulj2022stochastic} recently proposed for multipartite SBM. Research on how  how to choose and cheaply evaluate the weights in the SBM framework remains to be done. }

\section*{Acknowledgments}
The authors would like to thank St{\'e}phane Robin for his helpful advice.
This work was supported by a public grant as part of the Investissement d'avenir project, reference ANR-11-LABX-0056-LMH, LabEx LMH. This work was partially supported by the grant ANR-18-CE02-0010-01 of the French National Research Agency ANR (project EcoNet).

%========================================================

%========================================================
%========================================= BIBLIO
%=========================================================

\bibliographystyle{apalike}
\bibliography{bmpop_bib}

\pagebreak

\renewcommand{\thesection}{S-\arabic{section}}
\renewcommand{\theequation}{S-\arabic{equation}}
\renewcommand{\thefigure}{S-\arabic{figure}}
\renewcommand{\thetable}{S-\arabic{table}}
\renewcommand{\theremark}{S-\arabic{remark}}

\appendix

\section{Details of the network partitioning algorithm}\label{coll:ap:sec:partition}

\paragraph*{Definition of a dissimilarity measure between networks of a collection $(\Xm)_{m \in \Mcal}$}

It relies on the following steps.

\begin{enumerate}
 \item Infer   $\colSBM$ on $(X^{m})_{m \in \Mcal}$ to get coherent groupings of the nodes encoded in $\widehat{\btau}^{m} = (\widehat{\tau}^{m}_{iq})_{i=1,\dots,n_m,q=1, \dots, \widehat{Q}}$, for any  $m \in \Mcal$, where $\tau^m_{iq}$ is defined as in Section \ref{coll:sec:inference}.
  This step supplies a grouping of the nodes into blocks for each  network.% \new{enforcing common } mesoscale structures.
 Note that the inference also supplies the $(\widehat{\dens}^m)_{m \in \Mcal}$, these quantities being set to $1$ if we work with  the $\picolSBM$ and $\iidcolSBM$.
 \item For each network $m$,  compute:
 \begin{equation}\label{coll:eq:tilde}
 %\widetilde{n}^m_{qr}  = \sum_{\substack{i,j=1 \\ i\neq j} }^{\nm} \widehat{\tau}^{m}_{iq}\widehat{\tau}_{jr}^{m}, \quad
 \widetilde{\alpha}^m_{qr} = \frac{ \sum_{\substack{i,j=1 \\ i\neq j} }^{\nm} \widehat{\tau}^{m}_{iq}\widehat{\tau}^{m}_{jr} \Xm_{ij}}
 {\sum_{\substack{i,j=1 \\ i\neq j} }^{\nm} \widehat{\tau}^{m}_{iq}\widehat{\tau}_{jr}^{m}}, \quad \widetilde{\pi}^m_q = \frac{\sum_{i=1}^{\nm}\widehat{\tau}^m_{iq}}{\nm}
 \end{equation}%, \quad \widetilde{\delta}^m = \sum_{\substack{i,j=1 \\ i\neq j} }^{\nm} X_{ij}^m  $$
\label{coll:page:tilde}
 with the convention that $\widetilde{\pi}^m_q = 0$ if $q \notin \Qcal_m$ and $\alpha^m_{qr} = 0$ if $\{q,r\} \not\subset \Qcal_m$.
 These quantities are the separated estimates of the parameters encoding the mesoscale structure for each network, computed from nodes grouping obtained by considering all the networks jointly. \new{They correspond to the parameters estimates of the SBM when the block memberships are known. $\widetilde{pi}^m_q$ corresponds to the expected proportion of nodes of network $m$ grouped in block $q$, while $\alpha^m_{qr}$ is the expected connectivity parameter for network $m$ between group $q$ and $r$. }

 \item Then, for any pair of networks $(m,m') \in \Mcal^2$ compute the dissimilarity:
  \begin{equation}\label{coll:eq:dist_max}
      D_\Mcal(m, m')  =  \sum_{(q ,r)=1}^{Q} \max\left(\widetilde{\pi}^{m}_{q}, \widetilde{\pi}^{m'}_{q}\right)\max\left(\widetilde{\pi}^{m}_{r}, \widetilde{\pi}^{m'}_{r}\right) \left(\frac{\widetilde{\alpha}^{m}_{qr}}{\widehat{\dens}_m} - \frac{\widetilde{\alpha}^{m'}_{qr}}{\widehat{\dens}_{m'}}\right)^2.
    \end{equation}
\end{enumerate}

\new{The dissimilarity measure quantifies to what extent the connectivity parameters inferred separately on each network of the pair are different even though the nodes grouping were inferred jointly. This is weighted by the size of the blocks and corrected in the case of $\denscolSBM$ and  $\denspicolSBM$ by the density parameter. We take the maximum of the proportion of each block over the two networks in order to further increase the dissimilarity of networks not populating the same blocks. Two networks with low dissimilarity measure between them are expected to have more similar  mesoscale structure, and so are more expected to be part of the same sub-collection than two network with higher dissimilarity measure between them.}
%The enforced connectivity patterns of two networks with low dissimilarity measure between them have a  , it means that enforcing the same connectivity patterns by estimating common connectivity parameters for these two networks is not relevant and the networks cannot be considered to be part of the same sub-collection.}

%This dissimilarity measure quantifies to what extent the connectivity parameters inferred separately on each network of the pair are different.

%If this dissimilarity measure is large, it means that enforcing the same connectivity patterns by estimating common connectivity parameters for these two networks is not relevant and the networks cannot be considered to be part of the same sub-collection.

\paragraph*{An algorithm to cluster the collection of networks}

\new{Now, we use this dissimilarity to guide the search for the best partition of the collection of networks by using  Algorithm \ref{coll:algo:partition} which consists in a recursive partitioning of the collection. It relies on a $2$-medoids algorithm on the dissimilarity matrix defined in Equation \eqref{coll:eq:dist_max} to obtain a partition of the collection into two sub-collections of networks. This step is repeated recursively until the score, based on the $\BICL$, defined in Equation \eqref{coll:ICL:partition} calculated from the new partition stops increasing.}

\begin{algorithm}[!ht]\label{coll:algo:partition}
\textbf{Call:} \texttt{Clust2Coll($\obs=(\Xm)_{m\in \Mcal}$)}\\
\KwData{$\obs=(\Xm)_{m\in \Mcal}$ a collection of networks and $\mathcal{G}=\{\Mcal\}$ the trivial partition in a unique sub-collection}
\BlankLine
\Begin{
-Fit $\colSBM$  on $\obs$\\
-Compute the score $Sc_0 = Sc(\mathcal{G})$\\
-Compute the dissimilarity for all the networks in the collection $ \big(D_\Mcal(m,m')\big)_{m,m'\in\Mcal}$\\
% \SD{}{$m,m'\in \Mcal$??? }\\
-Apply a $2$-medoids algorithm to obtain $G_1$ and $G_2$ giving a partition of $\Mcal$.\\
-Compute $Sc^* = Sc(\mathcal{G}^*)$ where $\mathcal{G}^*=\{G_1,G_2\}$.
}
\BlankLine
\eIf{$Sc_0>Sc^*$}{\Return{$\mathcal{G}$}
}{\Return{\big\{\texttt{Clust2Coll}\big($(\Xm)_{m\in G_1}$\big),\texttt{Clust2Coll}\big($(\Xm)_{m\in G_2}$\big)\big\} }}
 %\Return{$\widehat{\Gcal} = \arg\max \Sc(\Gcal)$, $\widehat{\theta}$ and $\widehat{Z}$.}
\caption{Clustering a collection of networks into two sub-collections}
\end{algorithm}

\renewcommand{\thesection}{S-\arabic{section}}
\renewcommand{\theequation}{S-\arabic{equation}}
\renewcommand{\thefigure}{S-\arabic{figure}}
\renewcommand{\thetable}{S-\arabic{table}}
\renewcommand{\theremark}{S-\arabic{remark}}
\setcounter{properties}{0}
\setcounter{equation}{0}
\setcounter{figure}{0}
\setcounter{table}{0}
\setcounter{remark}{0}
\setcounter{section}{0}
%\titleformat{\section}{\large\bfseries}{\appendixname~\thesection .}{0.5em}{}

\newpage

{\Huge Supplementary Material}

%===============================================================================
\section{Proof of the identifiability of \colSBM s}\label{coll:sm:sec:identifiability_proof}
%===============================================================================
\begin{properties} \label{coll:prop:ident_colsbm}$\;$
    %The models are identifiable under the following asumptions and up to the following conditions:
  \begin{description}

    \item [$\iidcolSBM$] The parameters $ ( \bpi, \balpha)$  are identifiable up to a  label switching of the blocks provided that:
    %\textcolor{red}{A voir les preuves sans $\pi$ si on doit identifier les clusters des différents réseaux}
    \begin{enumerate}[label=(1.{\arabic*})]
      \item $\exists m^* \in \{1, \dots, M\} : n_{m^*} \geq 2Q$,
      %\item $\pi_{q} >0$ for all $q \in  \{1, \dots Q\}$
      \item $(\balpha \cdot \bpi)_{q} \neq (\balpha \cdot \bpi)_{r}$  $\forall (q,r) \in \{1, \dots, Q\}^2, q \neq r $.
    \end{enumerate}

    \vspace{1em}

  \item [$\denscolSBM$] The parameters $(\bpi, \balpha, \delta_1,\dots, \delta_M)$  are identifiable up to a  label switching of the blocks provided that:
  \begin{enumerate}[label=(2.{\arabic*})]
    \item $\exists m^* \in \{1, \dots, M\} : n_{m^*} \geq 2Q$ and $\dens_{m^*} = 1$,
    \item $(\balpha \cdot \bpi)_{q} \neq (\balpha \cdot \bpi)_{r}$  $\forall (q,r) \in \{1, \dots, Q\}^2, q \neq r $.

    \end{enumerate}

    \vspace{1em}

    \item [$\picolSBM$]  Assume that $ \forall m =1, \dots,M, \Xm \sim \Fcal \SBM_{\nm}(Q,\bpi^m,\balpha)$. Let $Q_m = |\Qcal_m| = |  \{q = 1\dots, Q, \pi^m_q >0\}|$ be  the number of non empty blocks in network $m$. Then the parameters  $(\bpi^1, \dots, \bpi^M, \balpha)$ are identifiable up to a  label switching of the blocks under the following conditions:
        \begin{enumerate}[label=(3.{\arabic*})]
        \item $\forall m \in \{1, \dots, M\}  : \nm \geq 2Q_m$,
        \item $\forall m \in \{1, \dots, M\}$, $(\balpha \cdot \bpim)_{q} \neq (\balpha \cdot \bpim)_{r}$  $\forall (q,r) \in \mathcal{Q}_m^2, q \neq r$,
        \item Each diagonal entry of $\balpha$ is unique.
    \end{enumerate}

        \vspace{1em}

     \item [$\denspicolSBM$]  Assume that $ \forall m =1, \dots,M, \Xm \sim \Fcal \SBM_{\nm}(Q,\bpi^m,\delta_m\balpha)$. Let $Q_m = |\Qcal_m| = |  \{q = 1\dots, Q, \pi^m_q >0\}|$ be  the number of non empty blocks in network $m$. Then the parameters  $(\bpi^1, \dots, \bpi^M, \balpha, \delta_1, \dots, \delta_M)$ are identifiable up to a  label switching of the blocks under the following conditions:

    \begin{enumerate}[label=(4.{\arabic*})]
        \item $\forall m \in \{1, \dots, M\}: \nm  \geq 2|\mathcal{Q}_m|$,
        \item $\dens_1 = 1$,
    \end{enumerate}
    \vspace{1em}

    \noindent If $Q \geq 2$:

    \begin{enumerate}
    \item[(4.3)] $(\balpha \cdot \bpim)_{q} \neq (\balpha \cdot \bpim)_{r}$ for all $(q \neq r) \in \mathcal{Q}_m^2$,
    \item[(4.4)] $\forall m \in \{1, \dots, M\}, Q_m \geq 2 $, %$\forall q \neq r = 1, \dots, Q, \quad \exists m : \{q, r\} \subset \Qcal_m$
    \item[(4.5)] Each diagonal entry of $\balpha$ is unique,
     \end{enumerate}
    \vspace{1em}

    \noindent If $Q \geq 3$:
%         \item There are no triplet $\{q, r, s\} \subset \Qcal$ such that $ \exists c \neq 0 : \con_{qq} = c\con_{rr} = c^2\con_{ss}$
%         \item There exists a permutation of the network indices  $\{1, \dots, M\}$, $\varsigma$ such that
%         \[\forall m \geq 2,  | \Qcal_{\varsigma(m)} \cap \cup_{l : \varsigma(l) < \varsigma(m)} \Qcal_{\varsigma(l)}| \geq 2.\]
%        \item Let $S$ be the graph with $Q$ nodes such that $q \leftrightarrow r$ if $\exists m : \{q, r\} \subset \Qcal_m$ of the two following condition is verified:
 %       \begin{enumerate}
  %        \item There exists a cycle of length $Q$
   %       \item
    %    \end{enumerate}
  \begin{enumerate}
    \item[(4.6)] There is no configuration of four indices $(q,r,s,t)\in\{1,\ldots,Q\}$ such that $\alpha_{qq}/\alpha_{rr}=\alpha_{ss}/\alpha_{tt}$ with $q\neq s$ or $r\neq t$ and with $q\neq r$ or $s\neq t$, % cases with r=t or q=r are disregarded because of assumption 5
\item[(4.7)] $\forall m \geq 2,  | \Qcal_{m} \cap \cup_{l : l < m} \Qcal_{l}| \geq 2.$
    \end{enumerate}
  \end{description}
\end{properties}

\begin{proof}

\citet{celisse2012consistency} proved that the parameters $(\bpim, \balpham)$ of the  $\Fcal \SBM_{\nm}(Q_m,\bpim, \balpham)$  are identifiable up to a label switching of the blocks  from the observation of a single network $X^m$ when $\Fcal$ is the Bernoulli distribution and provided that % networks $m \in \Mcal$  the parameters $\pim_{q}$ and $\con^m_{qr} = \dens_m\con_{qr}$  for all $q, r \in \Qcal_m$ up to label switching under the following assumptions:
 \begin{enumerate}
 \item $\nm > 2Q_m$,
 \item $(\balpham \cdot \bpim)_{q} \neq (\balpham \cdot \bpim)_{r}$ for all $(q \neq r) \in \{1, \dots, Q_m\}^2$.
% \item $\pim_q > 0$ and $\con^m_{qr} \in (0,1)$ for all $(q, r) \in \{1, \dots, Q_m\}^2$.
  \end{enumerate}
  Although they only consider the case where the emission distribution is the Bernoulli distribution, the extension to the Poisson distribution is
straightforward.
We prove the identifiability of our $\colSBM$ models by using their result,  $\Fcal$ being either the Bernoulli or the Poisson distribution.
The proofs for $\iidcolSBM$ and $\denscolSBM$ are straightforward while  the proofs for $\picolSBM$ and $\denspicolSBM$ are more complicated due to the possible existence of empty clusters in some networks $X^m$.
  \paragraph*{\underline{$\iidcolSBM$}}
Under this model,  for all $m=1, \dots M$, $\Xm \sim \Fcal\SBM_{\nm}(Q,\bpi, \balpha)$.
As a consequence, following \citet{celisse2012consistency}, the identifiability of  $\balpha$ and $\bpi$ is derived from the distribution of $X^{m^*}$ under assumptions $(1.1)$ and $(1.2)$.

\paragraph*{\underline{$\denscolSBM$}}
Under this model,   for all $m=1 \dots, M$,  $\Xm \sim \Fcal\SBM_{\nm}(Q,\bpi, \delta_m\balpha)$.
Under assumptions $(2.1)$ and $(2.2)$, we obtain the identifiability of   $\balpha$ and $\bpi$  from network $X^{m^*}$ \citep{celisse2012consistency}.
%Assuming that $\exists m^* : n_{m^*} \geq 2Q$ and $\delta_{m^*} = 1$,  we apply the theorem of \citet{celisse2012consistency} and obtain the identifiability of $\balpha$ and $\bpi$ under the condition $(\balpha \cdot \bpi)_{q} \neq (\balpha \cdot \bpi)_{r}$ for all $(q \neq r) \in \{1, \dots, Q\}^2$.
Now, for any $m \neq m^*$, by definition of $\Fcal\SBM_{\nm}(Q,\bpi, \delta_m\balpha)$, we have: $$\mathbb{E}[X^m_{ij}] = \delta_m \bpi' \balpha \bpi\,. $$
This proves that $ \delta_m$ is  identifiable.

\paragraph*{\underline{$\picolSBM$}}

Note that under $\picolSBM$, we have, for all $m$, $$\Xm \sim \Fcal\SBM_{\nm}(Q_m,\widetilde{\bpi}^m,\widetilde{\balpha}_m)$$  where $\widetilde{\bpi}^m$ is a vector of non-zero proportions of length $Q_m$ and $\widetilde{\balpha}^m$ is the restriction of $\balpha$ to $\Qcal_m$.
Under assumptions $(3.1)$ and $(3.2)$ and applying \citet{celisse2012consistency}  we obtain the identifiability of the parameters $\widetilde{\bpi}^m$ and $\widetilde{\balpha}^m$ from each network $\Xm$ separately.
However, the identifiability of each $\widetilde{\bpi}^m$ and $\widetilde{\balpha}^m$ is  established up to a label switching of the blocks in each network. We now have to find a coherent reordering between the networks which takes into account that some blocks are not represented in all the networks.

Let us build the complete matrix $\balpha$ using the $\widetilde{\balpha}^m$. We fill the diagonal of $\balpha$ which is composed of $(\mbox{diag}(\widetilde{\balpha}^m))_{m=1,\dots,M}$, taking the unique values and sorting them by increasing order, i.e.
$\alpha_{11} < \alpha_{22} < \dots < \alpha_{QQ}$.
This task is possible because of assumption (3.3).

Now, we get back to  $\widetilde{\bpi}^m$ and reorganize them to match with $\balpha$.  For any $m$, we define  $\phi_m: \{1, \dots, Q_ m\} \to \{1, \dots, Q\}$ such that  $\con_{\phi_m(q),\phi_m(q)} = \widetilde{\con}^{m}_{qq}$.
With the $(\phi_m)$ we are able to fill the rest of $\balpha$ as:
$\con_{\phi_m(q)\phi_m(r)} = \widetilde{\con}^{m}_{qr}$ for all $(q, r) \in \{1, \dots,Q_m \}^2$.
Finally, we define $\bpim$ a vector of size $Q$ such that: %$\pi^m_{q} = 0$ for any $q  \in \{1, \dots,Q\} \backslash \phi^m(\{1, \dots, Q_m\})$
\begin{equation*}
\pi^m_{q}  = \left\{
\begin{array}{cl}
  0   &  \forall q  \in \{1, \dots,Q\} \backslash \phi^m(\{1, \dots, Q_m\})\\
  \widetilde{\pi}^m_{ \phi_m^{-1}(q)}  & \forall  q \in  \phi^m(\{1, \dots, Q_m\})
 \end{array}
\right..
\end{equation*}

\paragraph*{\underline{$\denspicolSBM$}}
We now consider the model where
\begin{equation}\label{app:model deltapicol}
\Xm \sim \FSBM_{\nm}(Q, \bpim,\deltam   \balpha) .
\end{equation}
% We recall the assumptions.
% \begin{enumerate}
%         \item $\forall m  = 1, \dots M, : \nm  \geq 2Q_m$,
%         \item $\dens_1 = 1$,
%
%         For $Q \geq 2$:
%         \item $(\balpha \cdot \bpim)_{q} \neq (\balpha \cdot \bpim)_{r}$ for all $(q \neq r) \in \mathcal{Q}_m^2$,
%         \item $\forall m \in \{1, \dots, M\}, Q_m \geq 2 $, %$\forall q \neq r = 1, \dots, Q, \quad \exists m : \{q, r\} \subset \Qcal_m$
%         \item Each diagonal entry of $\balpha$ is unique,
%
%         For $Q \geq 3$:
% %         \item There are no triplet $\{q, r, s\} \subset \Qcal$ such that $ \exists c \neq 0 : \con_{qq} = c\con_{rr} = c^2\con_{ss}$
% %         \item There exists a permutation of $\{1, \dots, M\}$, $\varsigma$ such that
% %         \[\forall m \geq 2,  | \Qcal_{\varsigma(m)} \cap \cup_{l : \varsigma(l) < \varsigma(m)} \Qcal_{\varsigma(l)}| \geq 2.\]
% \item  There is no configuration of four indices $(q,r,s,t)\in\{1,\ldots,Q\}$ such that $\alpha_{qq}/\alpha_{rr}=\alpha_{ss}/\alpha_{tt}$ with $q\neq s$ or $r\neq t$ and with $q\neq r$ or $s\neq t$,
% \item $\forall m \geq 2,  | \Qcal_{m} \cap \cup_{l : l < m} \Qcal_{l}| \geq 2.$
% \end{enumerate}
%Assumptions 2 and 7 imply a specific order of networks. This could be extended but the proof becomes hard to follow
Like  in  model $\picolSBM$,  Equation \eqref{app:model deltapicol} implies that marginally, $$\Xm \sim \FSBM_{\nm}(Q_m, \widetilde{\bpi}^m,\widetilde{\balpha}^m) \quad  \mbox{ where }  \quad \widetilde{\balpha}^m = \delta_m (\alpha_{qr})_{q,r \in \Qcal_m} .$$
Applying \citet{celisse2012consistency} on the distribution of $\Xm$, we obtain the identifiability of the parameters $\widetilde{\bpi}^m$ and $\widetilde{\balpha}^m$ (assumptions (4.1) and (4.3)).  We now have to  match the structures of the networks and to take into account the empty blocks.
We separate the cases where $Q= 2$  from the ones where $Q >3$.

The proof relies on a sequential identification  the parameters (with respect to $m$).

\noindent $\bullet$ For $Q = 2$, we do not allow empty blocks (assumption (4.4), $Q_m \geq 2$) so  $ \widetilde{\balpha}^m =   \delta_m\balpha$ and $ \widetilde{\bpi}^m = \bpim$.
Using the fact that $\delta_1 = 1$ (assumption (4.2)), we identify $\bpi^1$ and $\balpha$.  Since we know that the diagonal entries of $\balpha$  are unique (assumption (4.5)), $\balpha$  can be chosen such that
$\con_{11} > \con_{22}$. This provides the ordering of the blocks in each network. Then, we identify the $\bpi^m$ in a unique manner and not up to label switching.
%and we identify the block permutation by the diagonal entries of $\balpha$ by noticing that as $\con_{11} \neq \con_{22}$ (for example , for each $m$ there is a unique permutation $\sigma_m$ such that $\widetilde{\balpha}^{m}_{\sigma_m(q)\sigma_m(q)} = \con_{qq}$ (such that  $\con^{m}_{\sigma_m(1)\sigma_m(1)} > \con^{m}_{\sigma_m(2)\sigma_m(2)}$) for all $q \in \{1, 2\}$.

\noindent $\bullet$   For $Q \geq 3$, for each $m \in \{1,\dots, M\}$, by assumptions $(4.1)$ and $(4.3)$ and using the marginal distributions,  we are able to identify $\widetilde{\balpha}^m$ and $\widetilde{\bpi}^m$.

\noindent Using the fact that $\delta_1 = 1$ (assumption (4.2)) and the fact that the entries of the diagonal of  $\balpha$ are unique,  we can do as in $\picolSBM$ and identify $\bpi^1$ and $(\alpha_{qr})_{(q, r) \in \Qcal_1}$.

%we can define the injection $\phi_1: \{1, \dots, |\Qcal_1| \} \to \Qcal$ and set $\con_{\phi_1(q)\phi_1(r)} = \con^{1}_{qr}$ for all $q, r \in \{1, \dots |\Qcal_1|\}$.
   Then for $m=2$, up to a relabelling of the blocks in $\widetilde\balpha^2$, we can define $\delta_2=\widetilde\alpha_{11}^2/\alpha_{11}=\widetilde\alpha_{22}^2/\alpha_{22}$ since there are at least two blocks in network $m=2$ that correspond to two blocks already identified in   network $m=1$ by assumption (4.7). We then need to prove that this parameter $\delta_2$ is uniquely defined.
   Let us assume that there exists a parameter $\delta_2'$ such that $\delta_2'\neq \delta_2$ and
   $\delta_2'=\widetilde\alpha^2_{ii}/\alpha_{kk}=\widetilde\alpha^2_{jj}/\alpha_{ll}$ with $i\neq j$. By definition of the $\denspicolSBM$
   $\alpha_{uu}:=\widetilde\alpha^2_{ii}/\delta_2$ which is either an already identified parameter from  network $m=1$ or corresponds to a new block  represented in network $m=2$ but not in network $m=1$.
   Note that, since $\delta_2'\neq \delta_2$,  then  $u\neq k$. For the same reason, $\exists v\neq l$ such that $\alpha_{vv}=\widetilde\alpha^2_{jj}/\delta_2$.
   Since $i\neq j$, the parameters $\widetilde\alpha^2_{ii}$ and
   $\widetilde\alpha^2_{jj}$ are not equal which implies that $\alpha_{kk}\neq\alpha_{ll}$ and $\alpha_{uu}\neq\alpha_{vv}$ and finally that $k\neq l$ and $u\neq v$.
   By computing:
   \begin{equation*}
    \frac{\alpha_{uu}}{\alpha_{vv}}=\frac{\widetilde\alpha^2_{ii}/\delta_2}{\widetilde\alpha^2_{jj}/\delta_2}\cdot \frac{\delta_2'}{\delta_2'}= \frac{\alpha_{kk}}{\alpha_{ll}}
   \end{equation*}
we obtain a contradiction with assumption (4.6). Therefore, $\delta_2=\delta'_2$.

 We can then identify the blocks in network $m=2$ by matching $\widetilde\alpha_{qq}^2/\delta_2$ with the $\alpha_{qq}$ already identified. The $\widetilde\alpha_{qq}^2/\delta_2$  that do not match with the previously identified parameters complete the matrix $\balpha$. The process is iterated with networks $m=3,\ldots,M$. Once the matrix $\balpha$ and the parameters in $\bdelta$ are identified, injections from $\{1,\ldots,Q_m\}\rightarrow\{1,\ldots,Q\}$, corresponding to the matching of the blocks, provide the $\bpi^m$.

  \end{proof}

  \section{Variational estimation of the parameters}\label{coll:sm:sec:inference}

We provide here some details for  the estimation of the parameters $\btheta_S \in \Theta_S$ for a given support matrix $S$. For ease of reading, the index $S$ is dropped in this section.
The likelihood
\begin{equation}\label{coll:eq:likelihood}
  \ell(\obs;\btheta) = \sum_{m=1}^M \log \int_{\Zm}\exp\left \{\ell(\Xm | \Zm; \balpha, \bdelta) + \ell(\Zm;\bpi)\right\}d\Zm
\end{equation}
is not tractable in practice, even for a small collection of networks as it relies on summing over $\sum_{m = 1}^M|{\Qcal_m}|^{\nm}$ terms.
A well-proven approach to handle this problem for the inference of the SBM is to rely on a variational version of the EM algorithm. This is done by maximizing a lower (variational) bound of the log-likelihood of the observed data \citep{daudin2008mixture}. %The approach is similar for both Bernoulli and Poisson models.
More precisely,

\begin{eqnarray*}\label{coll:eq:vbound1}
  \ell(\obs ; \btheta) & =  & \sum_{m =1} ^M \ell(\Xm ; \btheta)\\
  &\geq&   \sum_{m = 1}^M\Big(\ell(\Xm ; \btheta) - D_{\mbox{KL}}(\Rcal_m(\Zm)\|p(\Zm|\Xm; \btheta))\Big)
  \end{eqnarray*}
 where  $D_{\mbox{KL}}$ is the Kullback-Leibler divergence and $\Rcal_m$ stands for  any distribution on $ \Zm$.
 The last equation can be reformulated as:
\begin{eqnarray}\label{coll:eq:vbound}
  \ell(\obs ; \btheta) %& =  & \sum_{m =1}^M\E_{\Rcal}[\ell(\Xm ; \btheta)] - \E_q[\log(\Rcal(\Zm))] + \E_\Rcal[p(\Zm|\Xm)], \nonumber \\
  & \geq & \sum_{m = 1} ^M \Big( \E_{\Rcal_m}[\ell(\Xm, \Zm ; \btheta)] + \mathcal{H}(\Rcal_m)\Big) =:  \mathcal{J}(\Rcal, \btheta).
  \end{eqnarray}
 where $\Rcal=\otimes_{m=1}^M \Rcal_m$ and  $\mathcal{H}$ denotes the entropy of a distribution. Now, if, for all $m\in\{1,\ldots,M\}$, $\Rcal_m$
is chosen in the set of fully factorizable distributions and if  one sets $\taum_{iq} = \pr_{\Rcal_m}(\Zm_{iq} = 1)$ then $\mathcal{H}(\Rcal_m)$  is equal to:
\begin{equation}\label{coll:eq:entropy}
  \mathcal{H}(\Rcal_m) = - \sum_{i = 1}^{\nm}\sum_{q \in \Qcal_m} \taum_{iq} \log \taum_{iq}.
\end{equation}
Besides, the complete likelihood of network $m$  for the $\denspicolSBM$ marginalized over $\Rcal_m$ is given by:
\begin{equation}\label{coll:eq:comp_loglik}
  \mathbb{E}_{\mathcal{R}_{m}}[\ell(\Xm, \Zm ; \btheta)] = \sum_{\substack{i,j=1 \\ i\neq j} }^{\nm}\sum_{(q,r)  \in \Qcal_m} \taum_{iq}\taum_{jr} \log f(\Xm_{ij};\dens_m\con_{qr}) + \sum_{i = 1}^{\nm} \sum_{q \in \Qcal_m} \taum_{iq} \log \pim_{q}.
\end{equation}
Finally, the variational lower bound $ \mathcal{J}(\Rcal, \btheta) := \mathcal{J}(\btau, \btheta)$ is obtained by plugging Equations \eqref{coll:eq:entropy} and \eqref{coll:eq:comp_loglik} into the right member of Equation \eqref{coll:eq:vbound}. Note that the lower bound $\mathcal{J}(\btau, \btheta)$ is equal to the log-likelihood if $\Rcal_m(\Zm)  = p(\Zm|\Xm;\btheta)$ for all $m\in\{1,\ldots,M\}$.

The variational EM (VEM) algorithm  consists in optimizing the lower bound $\mathcal{J}(\btau, \btheta)$ with respect to $(\btau,\btheta)$, by iterating two optimization  steps with respect to $\btau$ and $\btheta$ respectively,
also referred to as VE-step and   M-step. The details of each step are specific to the  model at stake and are detailed hereafter.

\paragraph*{VE-step}
At iteration $(t)$ of the VEM algorithm, the VE-step consists in maximizing the lower bound with respect to $\btau$:
\begin{equation*}
  \widehat{\btau}^{(t+1)}= \arg \max_{\btau} \mathcal{J}(\btau, \widehat{\btheta}^{(t)}).
\end{equation*}
Note that by doing so, one minimizes the Kullback-Leibler divergences between $\Rcal_m(\Zm)$ and $p(\Zm|\Xm)$, and so approximates the true conditional distribution $p(\Zm|\Xm)$ in the space of fully factorizable probability distributions.
The $\btau^{m}$'s can be optimized separately by iterating the following fixed point systems for all $ m \in \{1, \dots, M\}$:
\begin{equation}\label{coll:eq:fixedpoint}
  \widehat{\tau}^{m(t+1)}_{iq} \propto \widehat{\pi}^{m(t)}_{q} \prod_{\substack{j=1 \\ j\neq i} }^{\nm}\prod_{r \in \Qcal_m}f(\Xm_{ij};\widehat{\dens}_{m}^{(t)}\widehat{\alpha}^{(t)}_{qr})^{\widehat{\tau}^{m(t+1)}_{jr}} \quad \forall i = 1, \dots, \nm, q \in \Qcal_m.
\end{equation}
% This equation has no explicit expression and  in practice the $(\widehat{\tau}^{m(t+1)}_{iq})$'s are obtained  by iterating   the fixed point Equation \eqref{coll:eq:fixedpoint}.

\paragraph*{M-Step}
At iteration $(t)$ of the VEM algorithm, the M-step   maximizes  the variational bound with respect to the model parameters $\btheta$:
\begin{equation*}
    \widehat{\btheta}^{(t+1)} = \arg \max_{\btheta} \mathcal{J}(\widehat{\btau}^{(t+1)}, \btheta ).
\end{equation*}
The update depends on the chosen model and the estimations are derived by canceling the gradient of the lower bound.  For the sake of simplicity, the iteration index $(t)$ is dropped  in the following formulae.
The obtained  formulae involve the following quantities:
\begin{eqnarray*}
  e_{qr}^{m} = \sum_{\substack{i,j=1 \\ i\neq j} }^{\nm} \taum_{iq}\taum_{jr} \Xm_{ij}, \quad
  \quad  n_{qr}^{m} = \sum_{\substack{i,j=1 \\ i\neq j} }^{\nm} \taum_{iq}\taum_{jr},  \quad n^{m}_{q} = \sum_{i=1}^{\nm} \taum_{iq}.
\end{eqnarray*}
% which represent respectively, the expected number of interactions  between blocks $q$ and $r$ in network $m$, the expected number of possible interactions and the average number of nodes in cluster $q$ of network $m$. Notice that if $q \neq r$, then $n^m_{qr} = n^m_{q}n^m_{r}$.\\
On the one hand, %, for the models with varying block proportions ($\picolSBM$ and $\dens\picolSBM$)
the $(\pi^{(m)}_q)_{q \in \Qcal_m}$ are estimated as%
$$\widehat{\pi}^m_q = \frac{n^m_q}{\nm}  \quad \quad \mbox{for $\picolSBM$ and $\dens\picolSBM$,}$$
which is the expected proportion of the nodes  in each allowed block  for network $m$.
On the other hand,  %for the models with non-varying block proportions ($\iidcolSBM$ and $\denscolSBM$) the proportions $\widehat{\pi}_q$ are computed :
$$\widehat{\pi}_q = \frac{\sum_{m=1}^M n^m_q}{\sum_{m=1}^M  \nm}  \quad \quad \mbox{for $\iidcolSBM$ and $\denscolSBM$}, $$
taking into account all the networks at the same time.
The connection parameters $\alpha_{qr}$  of $\iidcolSBM$ and $\picolSBM$
are estimated as the ratio of the number of  interactions between blocks $q$ and $r$  among all networks over the number of possible interactions:
$$\widehat{\con}_{qr} = \frac{\sum_{m=1}^M e^{m}_{qr}}{\sum_{m=1}^M n^{m}_{qr}} \quad  \mbox{for $\iidcolSBM$ and $\picolSBM$ }. $$
For the   $\denscolSBM$ and $\dens\picolSBM$,  there is no closed form for $\widehat{\bdelta}$ and $\widehat{\balpha}$ for a given value of $\btau$.
If $\Fcal = \mathcal{P}oisson$, then $\widehat{\bdelta}$ and $\widehat{\balpha}$ can be iteratively updated using the following formulae:
$$\hat{\con}_{qr} = \frac{\sum_{m=1}^M  e^{m}_{qr}}{\sum_{m=1}^M n^{m}_{qr} \hat{\dens}^{m}}\quad \mbox{ and } \quad  \hat{\dens}^{m} = \frac{\sum_{q,r \in \Qcal_{m}} e^{m}_{qr}}{\sum_{q,r \in \Qcal_m}  n^{m}_{qr} \hat{\con}_{qr}} $$
If $\Fcal = \mathcal{B}ernoulli$, no explicit expression can be derived and one has to rely  on a gradient ascent algorithm to update the parameters at each M-Step. %A closed-form formula exists for the Poisson model where $\dens$ and $\con$ depend on one another. So in this case we update the parameters  by iterating until convergence the update of $\dens$ and $\con$ given in Table \ref{coll:tab:mstep}.
%A summary of the formulas is provided in Table \ref{coll:tab:mstep},
\new{
\begin{remark}
  In practice when $\Fcal = \mathcal{B}ernoulli$, to update the parameters  $\widehat{\bdelta}$ and $\widehat{\balpha}$ we use by default the method of moving asymptotes \citep{svanberg2002class}. A much faster method consists in making a non constrained optimization by using the same estimates as the ones for $\Fcal = \mathcal{P}oisson$. The estimates $\widehat{\alpha_{qr}\delta_m}$ are clipped to $(0,1)$ for all $q,r,m$. Users of the \texttt{colSBM} package can choose which of these two methods to use as well as any of the ones implemented in the \texttt{NLOPT} package {\citep{nloptr}.}
\end{remark}
}
\begin{remark}
     In the VE-Step, each network can be treated independently, so the computation can be parallelized with ease.     Also, it can be more efficient to update only a subset of networks at each step to avoid being stuck in local maxima. So we use a slightly modified VEM algorithm where we just compute the VE-step on one network at a time (the  order of which is taken uniformly at random) before updating the corresponding parameters in the M-Step.
\end{remark}

%===============================================================================
\section{Details of the model selection procedure when allowing for empty blocks}\label{coll:sm:sec:detail_icl}
%===============================================================================

% \textcolor{red}{Propostion de changement pour le critère de sélection de modèle}

For $\picolSBM$ and $\dens\picolSBM$, the model is described by its support $S$.
 We can compute the likelihood for a given support. We recall that $\btheta_S = \{\balpha_S, \bdelta, \bpi_S \}$ are the parameters restricted to their support. %i.e. $\{(m,q) : S_{mq} = 1\}$ for $\bpi{|S}$ and $\{ (q,r) : (S'S)_{qr} > 0\}$ for $\balpha{|S}$.
Then for the model represented by $S$, the complete likelihood is given by:

\begin{eqnarray}\label{coll:eq:p_xz_given_s}
  p(\obs, \lat|S)& = & \int_{\btheta_S}p(\obs,\lat | \btheta_S, S) p(\btheta_S) \text{d}(\btheta_S) \nonumber \\
    & = &
  \int_{(\balpha_S,\bdelta)}\int_{\bpi_S} p(\obs | \lat, \balpha_S, \bdelta, S) p( \lat | \bpi_S, S)p(\balpha_S, \bdelta) p(\bpi_S)\text{d}(\balpha_S,\bdelta)\text{d}(\bpi_S) \nonumber \\
  & = & \nonumber
  \underbrace{\int_{(\balpha_S,\bdelta )}p(\obs | \lat, \balpha_S, \bdelta, S) p(\balpha_S, \bdelta )\text{d}(\balpha_S,\bdelta)}_{B1}  \underbrace{\int_{\bpi_S}p( \lat | \bpi_S, S) p(\bpi_S)\text{d}(\bpi_S)}_{B2},\\
\end{eqnarray}
where the prior on the  emission parameters and on the mixture parameters are assumed  to be independent.

The restriction of the parameter space to the one associated with the support $S$ is needed. Otherwise, some parameters would not be defined or would lie on the boundary of the parameters space, and the following asymptotic derivation would not be properly defined.
We use a BIC approximation on $B1$ where we rewrite:
\begin{eqnarray*}
  p(\obs|\lat, S) & = & \int_{(\balpha_S, \bdelta)} \left(\prod_{m =1}^{M} p(\Xm | \Zm, \balpha_S, \bdelta, S)\right) p(\balpha_S, \bdelta )\text{d}(\balpha_S,\bdelta) \\
  & = & \max_{(\balpha_S, \bdelta )}\exp \left(\sum_{m =1}^{M} \ell(\Xm | Z^m;\balpha_S, \bdelta,S )  -\frac{1}{2} \nu(\balpha_S,\bdelta)\log\left(\sum_m n_m(n_m-1)\right) + \mathcal{O}(1)\right),
\end{eqnarray*}
where
\begin{equation*}
 \nu(\balpha_S,\bdelta) =
\left\{
 \begin{array}{cl}
\nu(\balpha_S) = \sum_{q,r =1}^{Q} \mathbf{1}_{(S'S)_{qr}>0}, & \mbox{ for } \picolSBM\\
 \nu(\balpha_S)+M-1 & \mbox{ for } \denspicolSBM.
 \end{array}
\right.
\end{equation*}
 %And where we plugged in the maximum likelihood estimates of $\Zm$ and $(\balpha|S, \bdelta)$.
For $B2$, we use a $Q_m$-dimensional Dirichlet prior for each $\pi^m_S$:
\begin{eqnarray*}
  p(\lat|S) &= &\prod_{m\in \Mcal} \int_{\pi^m_S} p(\Zm|\pim_{S})p(\pim_S) \text{d}(\pim_S) \\
  & = & \max_{\bpi_S}\exp \left( \sum_{m=1}^{M} \ell(Z^m;\pi^{m}_S) - \frac{Q_m - 1}{2}\log(n_m)  + \mathcal{O}(1) \right).
\end{eqnarray*}
Then, we input $B1$ and $B2$ into Equation \eqref{coll:eq:p_xz_given_s} to obtain:
\begin{equation}\label{coll:eq:icl_noent}
  \log p(\obs,\lat|S)  \approx   \max_{\btheta_S} \ell(\obs, \lat;\btheta|S) -\frac12 \big(\pen_\pi(Q,S)+\pen_{\alpha}(Q,S)+ \pen_{\delta}(Q,S)\big),
\end{equation}
with the penalty terms as given in the main text.
% \begin{eqnarray*}\label{coll:eq:pen_Q}
%   \pen_\pi(Q,S)  &=& \sum_{m=1}^M\left(Q_m-1\right)\log(n_m),\\
% \pen_{\alpha}(Q,S) &=& \left(\sum_{q,r=1}^{Q}\mathbf{1}_{(S'S)_{qr} > 0}\right)  \log\left(N_M\right),\\
%  \pen_{\delta}(Q,S)  &=& \left\{
% \begin{array}{ll}
%  0 & \mbox{for $\picolSBM$}\\
% (M-1) \log \left(N_M\right) & \mbox{for $\denspicolSBM$}
% \end{array}
% \right. ,\\
%  \end{eqnarray*}
%
% \begin{equation*}
%    \mbox{pen}(S) = \sum_{m=1}^{M} \frac{Q_m - 1}{2}\log(\nm)  + \frac{1}{2} (\nu(\balpha|S) + \nu(\bdelta))\log\frac{\sum_m \nm(\nm-1)}{2}.
% \end{equation*}
%

\noindent  As $\lat$ is unknown we replace each $Z^m_{iq}$ by the variational parameters $\hat{\tau}^m_{iq}$ which maximizes the variational bound for a given support $S$.
Then,  we add the entropy of the variational distribution $\mathcal{H}(\mathcal{\widehat{\Rcal}})$ to Equation \eqref{coll:eq:icl_noent}. This leads to the variational bound of Equation \eqref{coll:eq:vbound}, as
\[ \max_{\btheta_S}\mathcal{J}(\hat{\btau}, \btheta_S) = \max_{\btheta_S} \ell(\obs, \E_{\widehat{\Rcal}}[\lat];\btheta_S) + \mathcal{H}(\widehat{\Rcal}), \]
which we recall is a surrogate of the log-likelihood of the observed data.
We define
$$\BICL(\obs,Q,S)=\max_{\btheta_S}\mathcal{J}(\hat{\btau}, \btheta | S)-\frac12 \big(\pen_\pi(Q,S)+\pen_{\alpha}(Q,S)+ \pen_{\delta}(Q,S)\big)$$
which is a penalized likelihood criterion when the support $S$ is known.
%
% Furthermore, the $\btheta_{|S}$ which maximizes Equation \eqref{coll:eq:icl_noent} corresponds to:
% \begin{equation*}
%     \widehat{\btheta}_{|S} = \arg \max_{\btheta_{|S}} \mathcal{J}(\widehat{\btau}, \btheta_{|S}).
% \end{equation*}

\noindent Finally to obtain the criterion $\BICL(\obs,Q)$ for $\picolSBM$ and $\denspicolSBM$, we need to penalize for the size of the space of possible models that depends on the support $S$.
For a given $Q$ corresponding to the number of different blocks in the collection of networks $\obs$,
we set the prior on $S$ decomposed as the product of uniform priors  on the numbers of blocks (between $1$ and $Q$) actually represented in each network and uniform priors for the choice of these $Q_m$ blocks among the $Q$ possible blocks ($Q_m$ is the number of  blocks that are represented in network $m$):
 $$p_Q(S)=  p_Q(Q_1,\ldots,Q_M)\cdot p_Q(S|Q_1,\ldots,Q_M)= \frac{1}{Q^M} \cdot \prod_{m=1}^M 1\bigg/{{Q \choose Q_m}}\,.$$
The prior is given on the space $\Scal_Q$ of admissible support.
% we introduce for a maximum number of blocks and a fixed sequence of number of blocks $\Qbf = (Q, Q_1, \dots, Q_M)$, the space of possible models $\Scal_{\Qbf}$. We would like to penalize its size
% \begin{equation}\label{coll:eq:param_space}
%   |\Scal_{\Qbf}| = \prod_{m = 1}^{M} {Q \choose Q_m}.
% \end{equation}

% Then, let $\xi(S)$ be a prior probability on $S$, which we choose to be uniform on $\Scal_{\Qbf}$. We aim at computing :
Using a BIC approximation and under a concentration assumption on the correct support, we derive
\begin{eqnarray*}
  \log p(\obs,\lat|\Qbf) & = & \log \int_{S} p(\obs,\lat|S, \Qbf)p_Q(S)\text{d}S \\
  & \approx & \log \int_{S} \exp\big(\BICL(\obs,Q,S)p_Q(S)\big)\text{d}S \\
    & \approx & \max_{S\in \Scal_{Q}}\big( \BICL(\obs,Q,S) - \log p_Q(S)\big)\,. \\
%     & = &\max_{S\in \Scal_{\Qbf}} BIC_{\colSBM}(S) - \mbox{pen}(\Qbf) =: eBIC_{\colSBM}\,
\end{eqnarray*}
% where we assumed some concentration results on $BIC_{\colSBM}(S)$. This seems reasonable as the identifiability results obtained in Proposition \ref{coll:prop:ident_colsbm} is  for a given  $(Q_1, \dots, Q_M)$, which is stronger than for a given support $S$.
Thus, by denoting $\pen_{S}(Q)=- 2 \log p_Q(S)$ in the equation above, we obtain:
\begin{equation*}
  \eBICL(\obs,Q)  =       \max_{S\in \Scal_Q} \left [ \max_{\btheta_S} \mathcal{J}(\widehat{\btau}, \btheta_S)- \frac{1}{2}\left[\pen_\pi(Q,S) + \pen_{\alpha}(Q,S) + \pen_{\delta}(Q,S)+ \pen_{S}(Q)\right] \right],
\end{equation*}
where
\begin{eqnarray*}\label{coll:eq:pen_Q}
  \pen_\pi(Q,S)  &=& \sum_{m=1}^M\left(Q_m-1\right)\log(n_m),\\
\pen_{\alpha}(Q,S) &=& \left(\sum_{q,r=1}^{Q}\mathbf{1}_{(S'S)_{qr} > 0}\right)  \log\left(N_M\right),\\
 \pen_{\delta}(Q,S)  &=& \left\{
\begin{array}{ll}
 0 & \mbox{for $\picolSBM$}\\
(M-1) \log \left(N_M\right) & \mbox{for $\denspicolSBM$}
\end{array}
\right. ,\\
\pen_{S}(Q)&=&- 2 \log p_Q(S).%\left(   M \log Q + \sum_{m=1}^M \log   {Q \choose \sum_{q=1}^Q S_{mq}} \right)
 \end{eqnarray*}

%===============================================================================
\section{Simulation studies}\label{coll:sm:sec:simulation}
%===============================================================================

In this section, we perform a large simulation study.
The first study  aims at testing the ability of the inference method to recover the number of blocks and the parameters for the $\picolSBM$ model.
The second study highlights the performances in terms of clustering of networks based on their mesoscale structure.

%In these numerical studies, we focus on the ability of our model selection criterion to recover the model from which the simulated data have been generated and of our method to recover the block memberships and the connectivity parameters. We also test our partitioning methods on their ability to group networks with common connectivity structures.

%In this section, for the inference of $(\dens\text{--}\dens\pi)\colSBM$s, we will use the Poisson model as we have a closed-form for the $M$-step for the $V\text{--}EM$ algorithm, in order to reduce the computational cost of the simulations. We also noticed that doing so sometimes held better results than using the Bernoulli model.

\subsection{Efficiency of the inference procedure}

\paragraph*{Simulation paradigm}
Let us simulate data under  the $\picolSBM$ model with  $M=2$, $\nm = 120$ and $Q = 4$.
$\balpha$ and $\bpi$ are chosen as:
\begin{equation}\label{coll:val param sim}
\balpha = .25 + \begin{pmatrix}
    3\epsilon_{\con} & 2\epsilon_{\con} & \epsilon_{\con} & -\epsilon_{\con}\\
    2\epsilon_{\con}  & 2\epsilon_{\con} & -\epsilon_{\con} & \epsilon_{\con} \\
    \epsilon_{\con}  & -\epsilon_{\con} & \epsilon_{\con} & 2\epsilon_{\con} \\
    -\epsilon_{\con}  & \epsilon_{\con} & 2\epsilon_{\con} & 0
\end{pmatrix},  \quad \bpi^{1} = \sigma_{1}(.2, .4, .4, 0), \quad \bpi^{2} = \sigma_{2}(0, \tfrac{1}{3},\tfrac{1}{3},\tfrac{1}{3}).
\end{equation}
with $\epsilon_{\con}$ taking eight equally spaced values ranging from $0$ to $0.24$.
For each value of $\epsilon_{\con}$,  $30$ datasets $(X^1, X^2)$ are simulated, resulting in $8\times 30 = 240$ datasets. More precisely, for each dataset, we pick  uniformly at random two  permutations  of $\{1,\dots,4\}$ $(\sigma_1, \sigma_2)$ with the constraint that $\sigma_1(4) \neq \sigma_{2}(1)$. This ensures that each of the two networks have a non-empty block that is empty in the other one.
Then the networks are simulated with $\mathcal{B}\text{ern}\SBM_{120}(4,\balpha,\bpi^m)$ with the previous parameters.

%and where  %(the permutations are applied to the columns of the support matrix $S = \begin{pmatrix} 1 & 1 & 1 & 0 \\ 0 & 1 & 1 & 1 \end{pmatrix}^t$ as well).
%We simulate $30$ times for each $\epsilon_{\con}$ a collection of $2$ networks with $100$ nodes each.

Each network has $2$ blocks in common and their connectivity structures encompass a mix of core-periphery, assortative community and disassortative community structures, depending on which $3$ of the $4$ blocks are selected for each network. $\epsilon_{\con}$ represents the strength of these structures, the larger, the easier it is to tell apart one block from another.

\paragraph*{Inference}

On each simulated dataset, we fit the $\iidcolSBM$,  $\picolSBM$  and  $\sepSBM$ models.  The inference is performed with the VEM algorithm and the $\BICL$ criterions  presented in the main manuscript.
%For each dataset, we put in competition $\iidcolSBM$ with $\picolSBM$ and $\sepSBM$ with $\picolSBM$ using the $\BICL$ criterion introduced in Subsection \ref{coll:subsec:consensus choice}.

\paragraph*{Quality indicators}
The assess the quality of the inference, we compute the following set of indicators for each simulated dataset.
\begin{itemize}
 \item  First, for each dataset, we put in competition $\picolSBM$ with $\sepSBM$ and $\iidcolSBM$ respectively. To do so, for each dataset,  we compute the $\BICL$ of each model $\picolSBM$ is preferred to $\sepSBM$ (resp. $\iidcolSBM$) if its $\BICL$ is greater.
 \item Secondly, when considering the $\picolSBM$, we compare $\widehat{Q}$ to its true value ($Q=4$).
 \item For $\picolSBM$ and $Q$ fixed to its true value ($Q=4$),  we evaluate the quality of recovery of the support matrix $S$ by calculating:
\begin{equation}\label{coll:AdS}
\text{Rec}(\widehat{S},S) = \max_{\sigma \in \mathfrak{S}_{4}}\mathbf{1}_{\{\forall q,m S_{mq} = \widehat{S}_{m\sigma(q)}\}}
\end{equation}
the greater the better.
\item
In order to evaluate the ability to recover the true connectivity parameter in the $\picolSBM$ model,  we compare $\widehat{\balpha}$ to its true value for the true number of blocks $Q=4$ through:
\begin{equation}\label{coll:eq:RMSE}
  \text{RMSE}(\widehat{\con}, \con) = \min_{\sigma \in \mathfrak{S}_{4}} \sqrt{\frac{1}{16}\sum_{1 \leq q, r \leq 4} (\widehat{\con}_{\sigma(q)\sigma(r)} - \con_{qr})^2},
\end{equation}
the $\sigma$ being there to correct the possible label switching of the blocks.
%\SD{}{C'était un min, j'ai mis un max?}

\item
Finally,  we judge the quality of our grouping of the nodes into blocks with the Adjusted Rand Index \citep[][ARI = $0$ for a random grouping and $1$ for a perfect recovery]{hubert1985comparing}.
For each network, for the $\picolSBM$, using $\hat{Q}$,  we compare the block memberships to the real ones by taking the  average over  the two networks
$$\overline{\ARI} = \tfrac{1}{2}\left(\ARI(\widehat{\bZ}^1, \bZ^1) + \ARI(\widehat{\bZ}^2, \bZ^2)\right)$$
and by computing it on the whole set of nodes $$\ARI_{1,2} = \ARI\left((\widehat{\bZ}^1,\widehat{\bZ}^2),  (\bZ^1, \bZ^2)\right). $$

\end{itemize}

All these quality indicators are averaged  among the $30$ simulated datasets.
The results are provided in Table \ref{coll:tab:alpha}. Each line corresponds to the $30$ datasets simulated with  a given value of $\epsilon_{\con}$.  The first columns concatenate the results of  the model comparison task. The following set of columns is about the selection of $Q$ and the estimation of $S$. The last columns supply the $\text{RMSE}$ on $\balpha$ and the $\ARI$. % when the number of blocks $Q$ is known.  \SD{}{}

%we use the root mean square error between the parameter used for the simulation $\alpha$ and its estimate when the true number of blocks $Q = 4$ is known :

 \paragraph*{Results}

%in terms of model selection and recovering of the blocks and the connectivity parameter in Table \ref{}.

For the model comparison,  when $\epsilon_{\con}$ is small ($\epsilon_{\con} \in [0, .04]$), the simulation model is close to the  Erd\H{o}s-Rényi network and it is very hard to find any structure beyond the one of a single block. As such, the $\iidcolSBM$ and $\picolSBM$ models are equivalent and $\iidcolSBM$ is preferred to  $\sepSBM$.

We observe a transition when  $\epsilon_{\con} = .08$ % and $\epsilon_{\con} = .12$
where we become  able to recover the true number of blocks $\widehat{Q} = 4$ and the support of the blocks given the true number of blocks. During this transition, the model selection criterion is about half of the time in favor of $\sepSBM$ i.e.  the model with no common connectivity structure between the networks.

From $\epsilon_{\con} = .16$, we recover the true number of blocks and their support most of the time and the common structure obtained by the $\picolSBM$ is found to be relevant. Note that when we are able to recover the true number of blocks, we are also able to recover their support almost every time.

For both the estimation of the parameters and the ARIs, the results mainly follow our ability to recover the true number of blocks, with the error of estimation of the parameters slowly decreasing from $\epsilon_{\con} = 0.12$.
$\overline{\ARI}$ goes to $1$ a bit faster than $\ARI_{1,2}$, denoting our ability to recover faster the real grouping of the nodes of each network than to match the blocks between the networks. This is directly linked with the detection of the true number of blocks and their support. Indeed, to get $\ARI_{1,2} = 1$, we need $\text{Rec}(\widehat{S},S) = 1$  while the effective block number for each network is of only $Q = 3$, meaning that even with the wrong selected model we can still reach $\overline{\ARI} = 1$.

{\small
  \begin{table}
 \begin{center}
 \resizebox{\textwidth}{!}{%
   \begin{tabular}{c|cc|cccc|cccc}
     \hline
      & \multicolumn{2}{c|}{Model comparison} & \multicolumn{4}{c|}{Estimation of $Q$ and $S$}& \multicolumn{3}{c}{Parameter \& Grouping accuracy } \\
      & \multicolumn{2}{c|}{($\picolSBM$ vs $\cdot$)} & \multicolumn{4}{c|}{ under $\picolSBM$}& \multicolumn{3}{c}{under $\picolSBM$ (mean $\pm$ sd)}\\
      \hline

    \rule{0pt}{14pt} $\epsilon_{\con}$ &  $\sepSBM$ & $\iidcolSBM$ &   $\mathbf{1}_{\widehat{Q} < 4}$ & $\mathbf{1}_{\widehat{Q} = 4^\star}$ & $ \mathbf{1}_{\widehat{Q} > 4}$ & $\text{Rec}(\hat{S},S)$ & RMSE($\widehat{\alpha}$,$\alpha$) & $\overline{\ARI}$ & $\ARI_{1,2}$\\
     \hline
     $0$   & 1   & 0  & 1  & 0   & 0   & 0     & $.1  \pm .002$& 0             & 0\\
     $.04$ &.83   & 0  & 1  & 0   & 0   & 0     & $.13 \pm .003$& 0             & 0\\
    $.08$  & .27 &.43 & .97& .03 & 0   & .03   & $.13 \pm .044$& $.42 \pm .3$ & $.25 \pm .21$\\
    $.12$  & .4 & .73 & .3 & .67  & .03   & .67    & $.1  \pm .076$& $.95 \pm .04$ & $.64 \pm .29$ \\
    $.16$  & .9  & 1  & 0  & .93 & .07 & .9   & $.03 \pm .04$ & $.99 \pm .01$ & $.97 \pm .1$ \\
    $.2$   & .93 & 1  & 0  & .97 & .03 & .97   & $.02 \pm .04$ & $1$            & $.99 \pm .06$ \\
    $.24$  & .97   & 1  & 0  & 1   & 0   & 1     & $.01 \pm .003$& $1$           & $1$ \\
    \hline
   \end{tabular}}\caption{\textbf{Accuracy of the inference for varying $\balpha$}. All the quality indicators are averaged over the $30$ simulated datasets.} %The number of cluster $Q$  and the similarity of the block support $\text{Rec}(\hat{S},S)$ with the true one are given for $\picolSBM$.}
   \label{coll:tab:alpha}.
   \end{center}
 \end{table}
 }
 \subsection{Capacity to distinguish  $\picolSBM$ from  $\iidcolSBM$}

 % Varying the mixture parameters}
 We  aim to understand how well we are able to differentiate  $\iidcolSBM$ from  $\picolSBM$ depending on the block proportions. To do so, we fix $\nm = 90$ and $Q = 3$%.$\balpha$ as in equation \eqref{coll:val param sim} with $\epsilon_{\con} = 0.16$
 and set $\balpha$ and $\bpi$ as follows:
 \begin{eqnarray*}
    \balpha = .25 + \begin{pmatrix}
    3\epsilon_{\con} & 2\epsilon_{\con} & \epsilon_{\con} \\
    2\epsilon_{\con}  & 2\epsilon_{\con} & -\epsilon_{\con} \\
    \epsilon_{\con}  & -\epsilon_{\con} & \epsilon_{\con}
\end{pmatrix},
   \bpi^{1} = \left(\frac{1}{3}, \frac{1}{3}, \frac{1}{3}\right) & \text{and} & \bpi^{2} = \sigma\left(\frac{1}{3}-\epsilon_{\pi},\frac{1}{3},\frac{1}{3}+\epsilon_{\pi}\right),
 \end{eqnarray*}
 with $\epsilon_{\con} = 0.16$  and $\epsilon_{\pi}$ taking $8$ values equally spaced in $[0, .28]$.
 $\sigma$ is a random permutation of the blocks. We simulate $30$ different collections for each value of $\epsilon_{\pi}$. %The results are shown in Table \ref{coll:tab:pi}.

 Here again, we put in competition $\picolSBM$ with $\iidcolSBM$ and  $\sepSBM$ and select a model if its $\BICL$ the greater than the two other ones.
 Then, for $\picolSBM$ we compare $\widehat{Q}$ to $3$ and evaluate our ability to recover $S$. The results are provided in Table \ref{coll:tab:pi}.

First notice that, since we chose  $\epsilon_{\pi} \ll \tfrac{1}{3}$, we do not simulate any empty block. As a consequence,  the inference of the model is quite easy and  we are able to recover the true number of blocks and the right support for the $\picolSBM$ model almost always. When $\epsilon_{\pi} = 0$,  $\bpi^1 = \bpi^2$  and  the model reduces to  $\iidcolSBM$. This remark explains why $\iidcolSBM$ is preferred to $\picolSBM$ when $\epsilon_{\pi} < .2$. As $\epsilon_{\pi}$ increases,  $\picolSBM$ gets more and more selected, highlighting our capacity to recover the simulated structure. %Finally for $\epsilon_{\pi} = .24$, the second network will have an empty block with high probability and so we fail to recover the true support about half of the time and in a few case, we will prefer to choose two separated SBM to model this collection.

 \begin{table}
 \begin{center}
 %\resizebox{\textwidth}{!}{%
   \begin{tabular}{c|ccc|cc}
   \hline
    & \multicolumn{3}{c|}{Model comparison} & \multicolumn{2}{c}{Estimation of $Q$ and $S$}\\
    & \multicolumn{3}{c|}{} & \multicolumn{2}{c}{ under $\picolSBM$}\\%& \multicolumn{3}{c}{under $\picolSBM$ (mean $\pm$ sd)}\\
      \hline

     $\epsilon_{\pi}$ & $\iidcolSBM$ & $\picolSBM$ & $\sepSBM$ &  $\mathbf{1}_{\widehat{Q} = 3}$ &  $\text{Rec}(\hat{S},S)$\\
      \hline

    $0$   & \textbf{.97}    & .03   & 0   & 1   & 1  \\
    $.04$ & .97    & \textbf{.03}   & 0   & 1   & 1   \\
    $.08$ &  1  & \textbf{0}  & 0   & 1   & 1   \\
    $.12$ &  .77  & \textbf{.23}  & 0   & 1   & 1   \\
    $.16$ &  .77 & \textbf{.23} & 0   & 1   & 1 \\
    $.2$  &  .5 & \textbf{.5} &  0  & 1   &  1  \\
    $.24$ &  .2 & \textbf{.8}  & 0 & 1 & 1 \\
    $.28$ &  .03 & \textbf{.97}  & 0 & 1 & 1 \\
    \hline
   \end{tabular}%}
   \caption{\textbf{Model selection for varying mixture parameters}. The number of blocks $\widehat{Q}$ is given for the $\picolSBM$. The similarity of the block support  to the true one $\text{Rec}(\hat{S},S)$ is given for $\picolSBM$ with  $Q = 4$.}\label{coll:tab:pi}
   \end{center}
 \end{table}

%
%  \PB{}{se concentrer sur des ARI avec les blocs de tous les réseaux regroupés}
%
% \textcolor{red}{On ne traite pas trop des $2$ themes suivants, mais j'ai l'impression que ce n'est pas tres grave. On a deja beaucou de simu + appli dans le papier}
% \paragraph{Varying density parameter} We now look at the behavior of our model selection criterion when we make the density of some of the networks of the collection vary.
%
%
%
% \paragraph{Recovering the connectivity parameter} On all the above situations, we also look at how we are able to recover the true $\con$. We depict the result in Picture .

\subsection{Partitioning a collection of networks} \label{coll:sm:sec:partition}
The third simulation experiment aims to illustrate our capacity to perform a partition of a collection of networks based on their structure, as presented in Section 6 and Appendix A.

\paragraph*{Simulation scenario}
For $\iidcolSBM$, $\picolSBM$ and $\denscolSBM$ and $\denspicolSBM$, we simulate $M = 9$ undirected networks with $75$ nodes and $Q = 3$ blocks. The block  proportions are chosen as follows:
$$\bpi^1 = (.2, .3, .5)$$
and for all $m =2, \dots, 9$
\begin{equation*}
  \bpi^m = \left\{
  \begin{array}{llccc}
 \bpi^1  &  \mbox{ for } & \iidcolSBM & \mbox{ and } &  \denscolSBM \\
\sigma_m(\bpi^1)  &  \mbox{ for } & \picolSBM  & \mbox{ and } &   \denspicolSBM  \\
 \end{array}
\right.
\end{equation*}
where $\sigma_m$ is a  permutation of $\{1,2,3\}$ proper to network $m$ and   $\sigma(\bpi) = (\pi_{\sigma(i)})_{i=1,\dots, 3}$.
The networks are divided into $3$ sub-collections of $3$ networks with connectivity parameters as follows:
{\small
\begin{eqnarray}\label{coll:eq:alpha_partition}
  \balpha^{\text{as}} = .3 +\begin{pmatrix}
                   \epsilon & - \frac{\epsilon}{2} & - \frac{\epsilon}{2} \\
                   - \frac{\epsilon}{2} & \epsilon & - \frac{\epsilon}{2} \\
                   - \frac{\epsilon}{2} & - \frac{\epsilon}{2} & \epsilon
                 \end{pmatrix},   &
  \balpha^{\text{cp}} = .3 + \begin{pmatrix}
                   \frac{3\epsilon}{2}& \epsilon & \frac{\epsilon}{2} \\
                   \epsilon &  \frac{\epsilon}{2} & 0 \\
                    \frac{\epsilon}{2} & 0 &  -\frac{\epsilon}{2}
                 \end{pmatrix},  &
   \balpha^{\text{dis}} = .3 + \begin{pmatrix}
                   - \frac{\epsilon}{2}& \epsilon & \epsilon\\
                   \epsilon &  - \frac{\epsilon}{2} & \epsilon \\
                   \epsilon  & \epsilon &  - \frac{\epsilon}{2}
                 \end{pmatrix},
\end{eqnarray}
}
with $\epsilon \in [.1,.4]$. $\balpha^{\text{as}}$ represents a classical assortative community structure, while $\balpha^{\text{cp}}$ is a layered core-periphery structure with block $2$ acting as a semi-core. Finally, $\balpha^{\text{dis}}$ is a disassortative community structure with stronger connections between blocks  than within blocks.
If $\epsilon = 0$, the three matrices are equal and the $9$ networks have the same connection structure. Increasing $\epsilon$ differentiates the $3$ sub-collections of networks.
For $\denscolSBM$ an $\denspicolSBM$,   we add  density parameters $\delta^1 = \delta^4 = \delta^7 = 1$,
$\delta^2 = \delta^5 = \delta^8 = 0.75$ and $\delta^3 = \delta^6 = \delta^9 = 0.5$.
%\in \{1, .75, .5\}$ which are all used once in each group of networks.
%\SD{}{Pas clair : $\delta^1 = 1$ $\delta^4 = 0.75$ $\delta^7 = 0.5$ ???}

Each of these configurations is simulated $30$ times. We apply the strategy exposed in Section 6 and Appendix A of the main manuscript and evaluate  the recovery of the simulated network partition.

%\SD{
%We cluster the networks with a binary classification tree using the dissimilarity matrix describes in Equation \eqref{coll:eq:dist_max}, where we stop when there is no improvement on the $\BICL$. For $(\pi\text{-}\dens\pi)\colSBM$s, we also use a partition of networks where all networks in the same group share the same blocks obtained by infering the whole collection of network with a given $\colSBM$, ie. $\hat{\Qcal}_m = \hat{\Qcal}_{m'} \forall m, m' \in \Mcal_g \forall g \in \Gcal$. For a given set of networks in the collection, the inference algorithm is rerun $3$ times to avoid an unlucky run due do the stochasticity of the algorithm.}{

 %Each situation is simulated $30$ times and

% First we would like to point out that the problem is quite difficult, especially for $(\dens\text{-}\dens\pi)\colSBM$s and for non community structures. As illustrated in Figure \ref{coll:fig:net_classif}-A, inferring a SBM on one network at a time underestimate the true number of clusters ($9$ for $3$ networks with the same structure) a lot of the time (blue violin plot), while inferring $\colSBM$s allows us to recover finer blocks (red violin plot), but not perfectly and this is particular true for core-periphery strucutre.

\paragraph*{Results}
We assess the quality of our procedure by comparing the obtained partition  of the collection of networks with the simulated one through the ARI index. As $\epsilon$ grows we are able to better differentiate the networks and do so almost perfectly on all $\colSBM$ setup. Note that Adding complexity  slightly deteriorates the results as we  recover the partition better for $\iidcolSBM$ and $\picolSBM$s than for $\denscolSBM$ and $\denspicolSBM$ .

\begin{figure}[!ht]
\centering
    \includegraphics[width = \textwidth]{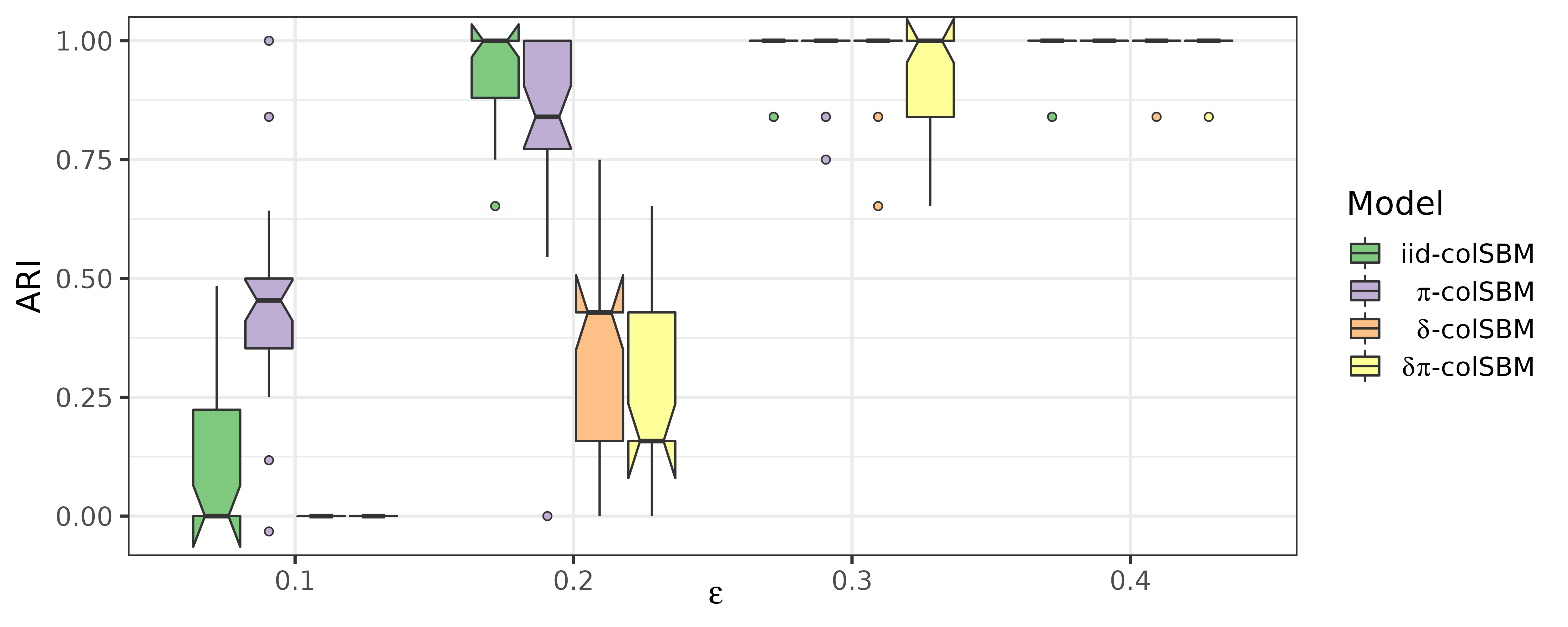}%\hfill% \includegraphics[width=0.3\textwidth]{collection/plot_classif_ari_dis_modif_legend}
  \caption{\textbf{Partition of networks}. ARI of the recovered partition of networks. %Orange is for the $\iidcolSBM$, green for the $\denscolSBM$, purple for the $\denspicolSBM$ and yellow for the $\picolSBM$.
  }\label{coll:fig:net_classif}% and the common block method (Support) when available (for ($\pi\text{-}\dens\pi)\colSBM$s).
\end{figure}
%}

\subsection{Finding finer block structures}
Finally, we perform a last simulation study in order to  illustrate the fact that, for particular configurations, using a  $\colSBM$  model on a collection of networks favors the transfer of information between networks and allows to  find finer block structures on  the networks. We consider the core-periphery structure  configuration described in Equation \eqref{coll:eq:alpha_partition} with $\epsilon = .4$. In that case $Q=3$.

We simulate a collection of $5$ networks. $4$ networks are of respective size $(90, 90, 120, 120)$.
The last network  is smaller with only  $60$ nodes and has a less marked structure ($\dens = .5$) for the  $\denscolSBM$ and $\denspicolSBM$ models.

Our goal is to recover the true connection structure of  this last network $X^5$.  To do so, we  compare the results obtained using either a standard single  $SBM$ on $X^5$, or using the  corresponding $\colSBM$ inferred with $M=2,3$ and $5$ networks.
We study $\widehat{Q}$ in the various scenarii. In the simulation experiment, we obtained only  $\widehat{Q}=2$ or $3$.  The experiment is repeated $30$ times.
The results are depicted in Figure \ref{coll:fig:net_blocks}.

For the  $4$ models of simulation, the simple $SBM$ recovers $2$ blocks most of the time. For $\iidcolSBM$ and $\picolSBM$, we always recover the $3$ blocks while for the other case, we improve the ability to recover the true number of blocks when the quantity of information available from the other networks grows, either by augmenting the number of networks or by augmenting the number of nodes.

% The simulation configuration is challenging and fitting an $SBM$ on each network fails to recover enough blocks some of the time, even for $\epsilon = .4$ as some blocks are quite small ($12$ expected nodes for the smallest one). To illustrate the interest of taking the information on this matter, for each structure, each model and each $\epsilon$, we look at the number of blocks recovered on each of the $3$ networks by the SBM and the one obtained by the corresponding $\colSBM$ using the information of the $2$ other networks with the same structure.

% For assortative and disassortative community structure, when $\epsilon \geq .3$, the adequate $(iid \text{--} \pi)\colSBM$ is always able to recover the true number of blocks, while the SBM sometimes fails to do so for $\epsilon = .3$. The task is harder for the configuration with the $(\dens \text{--} \dens\pi)\colSBM$s as the density is lower for some networks, but the number of blocks is always correctly estimated for $\epsilon = .4$, while the SBM fail to do so. The most challenging structure is the core-periphery structure with a semi-core, even for $\epsilon = .4$, in the  $(iid \text{--} \pi)\colSBM$ configuration, the $SBM$ finds only $2$ blocks about half of the time while the corresponding $\colSBM$ finds $3$ blocks constantly.
% The configurations involving $\dens$ are too challenging for both the $SBM$ and the $\colSBM$ for this number of nodes and while the $SBM$ are sometimes able to find $3$ blocks for $\epsilon = .4$ on the denser networks, the $\colSBM$ always fail to do so.

{\centering{
\begin{figure}[!ht]
    \includegraphics[width = \textwidth]{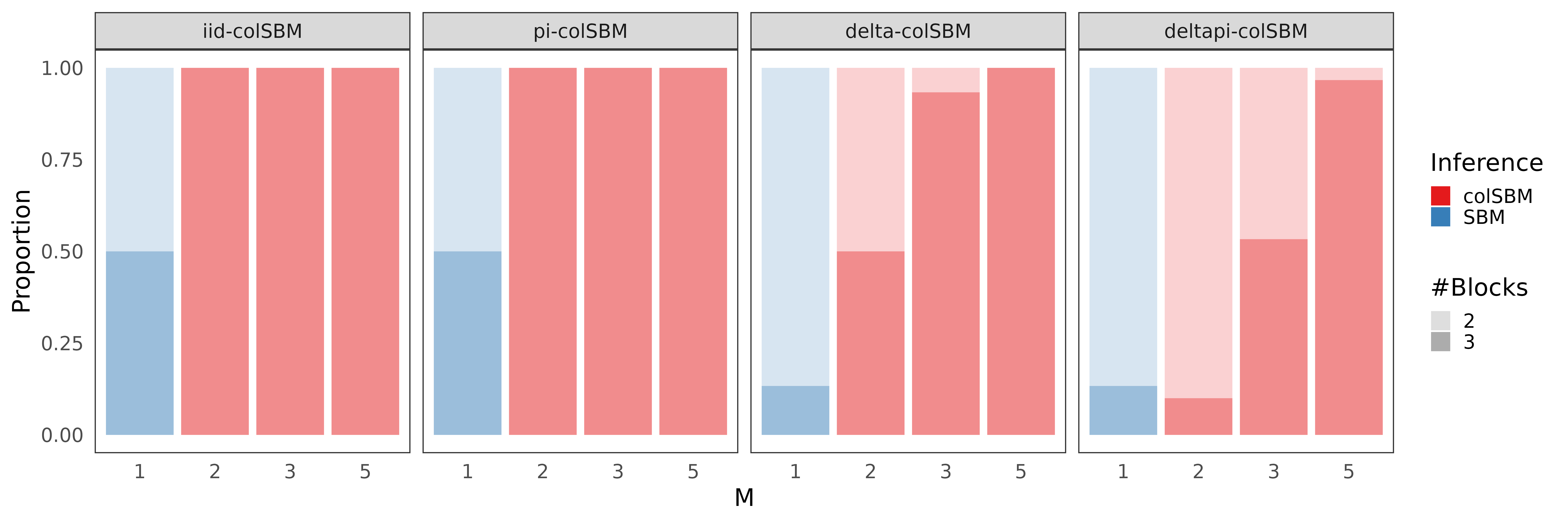}
  \caption{\textbf{Finding finer block structures.} Cumulative barplot of $\widehat{Q}$ by the $SBM$s (blue) and the adequate $\colSBM$ (red) under the different simulation scenario. The number of blocks to be recovered is $3$ and the darkest shade corresponds to $\widehat{Q} = 3$. }%of the corresponding color.}
\label{coll:fig:net_blocks}
\end{figure}
}}

\subsection{\new{A note on model selection: dealing with networks of different sizes}}

\new{We perform a simulation study to illustrate the fact that practitioner should be careful when dealing with networks of different sizes and should rely on $\BICL$  to state on the relevance of the common connectivity structure. We simulate $2$ directed binary networks using $\sepSBM$ with the following parameters:
\begin{eqnarray}\label{coll:eq:alpha_size}
  \balpha^{\text{as}} = \begin{pmatrix}
                   .55 & .1 & .1 \\
                   .1 & .5 & .1 \\
                   .1 & .1 & .45
                 \end{pmatrix},   &
  \bpi^{\text{as}} = (.4,.3,.3),
  \balpha^{\text{er}} = \begin{pmatrix}
                   .25
                 \end{pmatrix},  &
  \bpi^{\text{er}} = (1).
\end{eqnarray}
$X^{\text{as}} \sim \mbox{SBM}_{64}(3, \bpi^{\text{as}}, \balpha^{\text{as}})$ is drawn from an assortative community structure, while $X^{\text{er}} \sim \mbox{SBM}_{n_{\text{er}}}(1,\bpi^{\text{er}}, \balpha^{\text{er}})$ is an Erd\H{o}s-R\'{e}nyi network with $n_{\text{er}}$ ranging from $10$ to $640$.
The collection $(X^{\text{as}}, X^{\text{er}})$ is simulated $20$ times for each value of $n_{\text{er}}$ and inference is done for each $\colSBM$.
Our objective is to check if $\colSBM$ detects spurious structure on $X^{\text{er}}$ when $n_{\text{er}}$ is small and blurs the structure of $X^{\text{as}}$ when $n_{\text{er}}$ is large. We show the recovery of the grouping of the nodes into blocks for each network ($\ARI$) and the difference in $\BICL$ with the best model ($\sepSBM$) in Table \ref{coll:tab:size}.
}

 \begin{table}
 \begin{center}
 %\resizebox{\textwidth}{!}{%
   \begin{tabular}{c|llr|llr|llr|llr}
   \hline
    & \multicolumn{3}{c|}{$\iidcolSBM$} & \multicolumn{3}{c|}{$\picolSBM$} & \multicolumn{3}{c|}{$\denscolSBM$} & \multicolumn{3}{c}{$\denspicolSBM$}\\
      \hline
      & \multicolumn{2}{c}{$\ARI$} & & \multicolumn{2}{c}{$\ARI$} & & \multicolumn{2}{c}{$\ARI$} & & \multicolumn{2}{c}{$\ARI$} & \\
    $n_{\text{er}}$ & as  & er & $\Delta_{\BICL}$ & as  &  er  & $\Delta_{\BICL}$ & as & er & $\Delta_{\BICL}$ & as & er & $\Delta_{\BICL}$ \\
    \hline
     $10$ & $1$ & $0$ & $-5$          & $1$ & $.45$ & $-7$        & $1$ & $0$ & $-9$        & $1$ & $.9$ & $-6$ \\
     $20$ & $1$ & $0$ & $-16$         & $1$ & $.35$ & $-16$       & $1$ & $.15$ & $-17$     & $1$ & $.75$ & $-9$ \\
     $40$ & $.97$ & $.1$ & $-60$      & $1$ & $.45$ & $-19$       & $.99$ & $.45$ & $-41$   & $1$ & $.95$ &$-9$ \\
     $80$ & $.94$ & $.4$ & $-95$      & $1$ & $.95$ & $-14$       & $1$ & $.9$ & $-45$      & $1$ & $1$ & $-8$ \\
     $160$ & $.95$ & $.9$ & $-115$    & $.99$ & $1$ & $-16$       & $.99$ & $1$ & $-71$     & $1$ & $1$ & $-12$ \\
     $320$ & $.9$ & $1$ & $-141$      & $1$ & $1$ & $-24$         & $1$ & $1$ & $-95$       & $1$ & $1$ & $-18$ \\
     $640$ & $.7$ & $1$ & $-171$      & $.99$ & $1$ & $-26$       & $1$ & $1$ & $-132$      & $1$ & $1$ & $-24$ \\
   \end{tabular}%}
   \caption{Average recovery of the simulated block memberships ($\ARI$) for the assortative community networks (as) and the Erd\H{o}s-R\'{e}nyi networks (er) for each  $\colSBM$ and the average difference in $\BICL$ between $\colSBM$ and  the one used for the simulation ($\sepSBM$).}\label{coll:tab:size}
   \end{center}
 \end{table}

\new{In this setting, $\sepSBM$ always detects the correct structure for all networks and no $\colSBM$ is ever selected by the $\BICL$ criterion. When $n_{\text{er}}$ is small, $\iidcolSBM$, $\picolSBM$ and $\denscolSBM$ spuriously detect some structure on $X^{\text{er}}$ (low $\ARI$), while $\denspicolSBM$ is designed to correctly assign all the nodes of $X^{\text{er}}$ in one block.  As $n_{\text{er}}$ gets larger, the absence of structure on $X^{\text{er}}$ is correctly detected by all models. When $n_{\text{er}}$ is much larger than $n_{\text{as}}$, the structure of $X^{\text{as}}$ is blurred by the one $X^{\text{er}}$ while using $\iidcolSBM$ for the inference. %following occurs, for $\iidcolSBM$, most of the nodes of $X^{\text{er}}$ tend to concentrate on juste one block and blocks of $X^{\text{as}}$ merge, while the difference between $\BICL_{\iidcolSBM}$ and $\BICL_{\sepSBM}$ increases. For $\denscolSBM$, the concentration phenomenon is also observed for $X^{\text{er}}$, but the structure of $X^1$ is correctly recovered. $\picolSBM$ separates the blocks belonging to $X^{\text{er}}$ and the one to $X^{\text{as}}$ leading to $\hat{Q} = 4$ and a fully separated support. Finally $\denspicolSBM$ find the correct support for $X^{\text{er}}$ and merge it with a block of $X^{\text{as}}$.
The difference in $\BICL$ do not increase as fast for $\picolSBM$ and $\denspicolSBM$ compared to $\iidcolSBM$ and $\denspicolSBM$.}

%% if your bibliography is in bibtex format, uncomment commands:
     % Bibliography file (usually '*.bib')

%% or include bibliography directly:
% \begin{thebibliography}{}
% \bibitem[\protect\citeauthoryear{???}{???}]{b1}
% \end{thebibliography}

\end{document}